%% file: ChemPotential.tex
\documentclass[aps, prc, reprint, onecolumn, amsmath, amssymb, superscriptaddress, floatfix, nofootinbib, notitlepage]{revtex4-1} 

\usepackage[pdftex]{graphicx}
\DeclareGraphicsExtensions{.jpg,.png,.pdf}

\usepackage[utf8]{inputenc}
\usepackage{slashed}
\usepackage{hyperref}
\usepackage{amsmath}
\usepackage{amssymb}
\usepackage{amsfonts}
\usepackage{tabularx}
\usepackage{booktabs}
\usepackage{graphicx}
\usepackage{color}
\usepackage{multirow}
\usepackage[perpage,symbol]{footmisc}
\usepackage{verbatim}
\usepackage{dcolumn}
\usepackage{bm}
\usepackage[inline]{enumitem}
\definecolor{theblue}{RGB}{0,50,230}
\hypersetup{
  colorlinks=true,
  linkcolor=theblue,
  citecolor=theblue,
  urlcolor=theblue
}

\usepackage{lineno}

\begin{document}
	
\title{Soft-hard factorization of heavy-quark transport in QCD matter at finite chemical potential}
	
\author{Jiale~Lou}
\thanks{These authors contributed equally to this work.}%
\affiliation{%
	College of Mathematics and Physics, College of Nuclear Energy Science and Engineering, China Three Gorges University, Yichang 443002, China\\
}%

\author{Wu~Wang}
\thanks{These authors contributed equally to this work.}%
\affiliation{%
	College of Mathematics and Physics, College of Nuclear Energy Science and Engineering, China Three Gorges University, Yichang 443002, China\\
}%
	
\author{Jiazhen~Peng}
\affiliation{%
	College of Mathematics and Physics, College of Nuclear Energy Science and Engineering, China Three Gorges University, Yichang 443002, China\\
}%
	
\author{Fei~Sun}
\affiliation{%
	College of Mathematics and Physics, College of Nuclear Energy Science and Engineering, China Three Gorges University, Yichang 443002, China\\
}%
\affiliation{%
	Center for Astronomy and Space Sciences, China Three Gorges University, Yichang 443002, China\\
}%

\author{Kejun~Wu}
\affiliation{%
	College of Mathematics and Physics, College of Nuclear Energy Science and Engineering, China Three Gorges University, Yichang 443002, China\\
}%
\affiliation{%
	Center for Astronomy and Space Sciences, China Three Gorges University, Yichang 443002, China\\
}%
                                                                        
\author{Wei~Xie}
\affiliation{%
	College of Mathematics and Physics, College of Nuclear Energy Science and Engineering, China Three Gorges University, Yichang 443002, China\\
}%
\affiliation{%
	Center for Astronomy and Space Sciences, China Three Gorges University, Yichang 443002, China\\
}%

\author{Zuman~Zhang}
\affiliation{%
        School of Physics and Mechanical Electrical $\&$ Engineering, Hubei University of Education, Wuhan 430205, China\\
}%
\affiliation{%
	Institute of Theoretical Physics, Hubei University of Education, Wuhan 430205, China\\
}%

\author{Shuang~Li}
\email[Corresponding auther: ]{lish@ctgu.edu.cn}
\affiliation{%
	College of Mathematics and Physics, College of Nuclear Energy Science and Engineering, China Three Gorges University, Yichang 443002, China\\
}%
\affiliation{%
	Center for Astronomy and Space Sciences, China Three Gorges University, Yichang 443002, China\\
}%

\author{Sa~Wang}
\email[Corresponding auther: ]{wangsa@ctgu.edu.cn}
\affiliation{%
	College of Mathematics and Physics, College of Nuclear Energy Science and Engineering, China Three Gorges University, Yichang 443002, China\\
}%
\affiliation{%
	Center for Astronomy and Space Sciences, China Three Gorges University, Yichang 443002, China\\
}%

\date{\today}

\begin{abstract}
We calculate the collisional energy loss and momentum diffusion coefficients of
heavy quarks traversing a hot and dense QCD medium at finite quark chemical potential, $\mu\neq0$.
The analysis is performed within an extended soft-hard factorization model
that consistently incorporates the $\mu$-dependence of the Debye screening mass $M_D(\mu)$ and
of the fermionic thermal distribution functions.
Both the energy loss and the diffusion coefficients are found to increase with $\mu$,
with the enhancement being most pronounced at low temperatures
where the chemical potential effects dominate the medium response.
To elucidate the origin of this dependence,
we derive analytic high-energy approximations
in which the leading $\mu$-corrections appear as logarithmic terms:
a soft logarithm $\sim\mu^{2}\ln(|t^{*}|/M_{D}^{2})$ from $t$-channel
scattering off thermal gluonic excitations,
and a hard logarithm $\sim\mu^{2}\ln(E_{1}T/|t^{*}|)$ from scattering off thermal quarks.
In the complete result the dependence on the intermediate separation scale $t^{\ast}$ cancels, as required.
We also confirm the expected mass hierarchy $-dE/dz(charm)<-dE/dz(bottom)$ at fixed velocity.
Our findings demonstrate that finite chemical potential plays a significant role
in heavy-quark transport and must be included in theoretical descriptions of heavy-flavor dynamics
in baryon-rich environments, such as those probed in the RHIC Beam Energy Scan, and at FAIR and NICA.
\end{abstract}
	
	
\maketitle
	
\section{INTRODUCTION}\label{sec:Intro}
The quark-gluon plasma (QGP) is a deconfined state of strongly interacting matter
that existed in the early universe and
can be recreated in high-energy heavy-ion collisions~\cite{Gyulassy:2004zy, Shuryak:2004cy, STAR05, Frawley:2008kk,Bzdak:2019pkr}.
Experiments at the Relativistic Heavy Ion Collider (RHIC) and the Large Hadron Collider (LHC)
provide compelling evidence for QGP formation and
offer controlled environment to study its transport properties~\cite{Dong:2019byy, Tang:2020ame, Chen:2024aom, ALICE:2021rxa, ALICE:2022wpn}.
The macroscopic state of the QGP is characterized by its temperature $T$,
energy density, and conserved charges,
the latter being parametrized by chemical potentials.
In particular, the quark (or baryon) chemical potential $\mu$ governs
the net quark density and is especially relevant at moderate collision energies,
such in the RHIC Beam Energy Scan (BES), and at upcoming FAIR and NICA facilities~\cite{Chen:2024aom, CBM:2016kpk, MPD:2022qhn}.
A paramount goal of these experimental programs is to map the QCD phase diagram and to search for its conjectured critical endpoint (CEP)~\cite{Stephanov:1998dy}. Since the quark chemical potential $\mu$ acts as the control parameter driving the system toward criticality~\cite{Fodor:2004nz}, understanding its quantitative impact on heavy-quark transport in the normal (noncritical) phase is indispensable. Our present calculation therefore provides a well-defined baseline for identifying and isolating the possible enhancement of transport coefficients and energy loss induced by critical fluctuations in future BES-II, FAIR, and NICA data.

Heavy quarks (charm and bottom) are predominantly produced in initial hard scatterings and
subsequently propagate through the evolving medium.
Owing to their masses $m_Q\gg T$, they serve as calibrated probes:
their collisional energy loss and momentum diffusion encode information about microscopic scattering processes and
medium screening~\cite{RalfSummary16, Chen:2021akx, He:2022ywp}.
While extensive theoretical work has been devoted to heavy-quark transport at vanishing chemical potential,
most existing formulations assume $\mu=0$ and thus omit baryon-density
effects~\cite{PhysRevC.71.064904, Alberico:2011zy, PhysRevLett.110.112301, PhysRevC.93.034906, PhysRevC.92.024907, Cao:2018ews, CUJET3Arxiv18, CUJET3CPC18, PhysRevC.98.064901, CTGUHybrid1, Li:2019wri}.
A systematic extension to finite $\mu$ is essential for a realistic description of
heavy-flavor dynamics in baryon-rich environments.

Theoretical studies of collisional energy loss date back to Bjorken’s seminal estimate~\cite{bjorken1982energy},
which highlighted the role of multiple elastic scatterings.
Subsequent developments addressed the infrared sensitivity of the perturbative treatment.
Thoma and Gyulassy introduced resummation techniques for soft contributions~\cite{Thoma:1990fm},
while Braaten and Thoma established the soft-hard separation with resummed propagators to remove cutoff dependence~\cite{braaten1991energy}.
Later works have refined transport calculations using both perturbative and nonperturbative methods,
with some attempts to incorporate finite chemical potential~\cite{Vija:1994is, Berrehrah:2014tva, Jamal:2020emj, Madni:2022bea}.
Nevertheless, most analytical frameworks and phenomenological models still adopt the simplifying assumption $\mu=0$.

In this work, we extend the soft-hard factorization model (SHFM)~\cite{peng2024unraveling, li2021langevin}
to consistently include finite chemical potential.
The extension accounts for two key effects:
(i) the $\mu$-dependence of the Debye screening mass $M_D(T,\mu)$ in the soft sector,
and (ii) the modification of fermionic distribution functions that govern hard scattering rates.
Within this framework we compute the collisional energy loss $-dE/dz$ and
momentum diffusion coefficients $\kappa_{T/L}$ of heavy quarks across a range of $(E,T,\mu)$ values.
In addition to full numerical results, we derive analytic high-energy approximations
that isolate the leading $\mu$-dependent contributions in different scattering channels.
We explicitly verify the cancellation of the auxiliary separation scale $t^{*}$ between
soft and hard contributions, ensuring the consistency of the formalism.

The remainder of this paper is organized as follows.
In Sec.~\ref{sec:formalism} we summarize the extended SHFM formalism at finite chemical potential and define the relevant transport observables.
Section~\ref{sec:HEA} presents analytic high-energy approximations and discusses the origin and cancellation of the scale dependence.
Section~\ref{sec:results} contains the numerical results and phenomenological discussion,
focusing on the role of $\mu$ in baryon-rich QCD matter.
We conclude in Sec.~\ref{sec:conclusions} with a summary of the main findings and an outlook for future work.

Four-momenta are written $P^{\mu}=(p^{0},\vec{p}\;)$ and three-vectors as $\vec{p}$.
We use natural units $\hbar=c=1$ and the metric
$g^{\mu\nu}=\mathrm{diag}(1,-1,-1,-1)$ throughout the manuscript.

\section{Formalism}\label{sec:formalism}

\subsection{Soft-hard factorization model at finite chemical potential}\label{sec:gamma_HTLpQCD}
The elastic scattering process between a heavy quark and
a thermalized medium parton (light quarks $q$, antiquarks $\bar{q}$, or gluon $g$) is described by
\begin{equation*}
Q(P_1)+i(P_2)\to Q(P_3)+i(P_4),
\end{equation*}
where $P_{1}=(E_{1},\vec{p}_{1})$ and $P_2$ are the four-momenta of the incoming heavy quark ($Q$)
and medium parton ($i = q, \bar{q}, g$), respectively,
and $P_3$ and $P_4$ are the four-momenta of the outgoing particles.
The tree-level Feynman diagrams for the $t$-, $s$- and $u$-channels
can be found in Refs.~\cite{li2021langevin, peng2024unraveling},
together with the corresponding matrix elements.
The four-momentum transfer in each channel is
\begin{equation*}
Q^{\mu} = P^{\mu}_1-P^{\mu}_3=(\omega,\vec{q}\;)=(\omega,\vec{q}_T,q_L).
\end{equation*}

In the limit of small momentum transfer $t\rightarrow 0$,                                                                                                     
the $t$-channel gluon propagator induces an infrared divergence in the cross section, $d\sigma/dt\propto 1/t^2$.
To regularize this divergence, one must include contributions from long-wavelength gluons
with momentum scale $\sqrt{-t}\sim gT$, i.e., the soft scattering regime.
These soft exchanges are longrange in nature and
are screened by the medium through the Debye mass, necessitating the use of the
hard thermal loop (HTL) resummed gluon propagator~\cite{braaten1991calculation, PhysRevD.44.R2625, JeanPR02}.
In contrast, large momentum transfers $\sqrt{-t}\gtrsim T$,
i.e., the hard scattering regime, can be treated within perturbative QCD using the Born approximation.
The SHFM separates these two kinematic regions by introducing an intermediate cutoff $t^{\ast}$,
chosen such that $m_D^2 \ll -t^{\ast} \ll T^2$ at vanishing chemical potential,
which corresponds to the weak-coupling (or high-temperature) limit~\cite{braaten1991calculation, Romatschke:2004au, PBGPRC08}.
Physically, $|t^{\ast}|$ is of order the Debye screening scale, ensuring that
soft contributions are consistently resummed while hard ones remain perturbative.

Within this framework, the scattering rate receives contributions from three kinematic regions:
\begin{enumerate}
	\item[(1)] Soft region ($-t < -t^{\ast}$):
	heavy quark scattering off medium partons via $t$-channel exchange with small momentum transfer;
	the interference with $s/u$-channels is neglected;
	
	\item[(2)] Hard region ($-t > -t^{\ast}$):
	heavy quark scattering off medium partons via $t$-channel exchange with large momentum transfer;
	the interference with $s/u$-channels is included and attributed to the $t$-channel;
	
	\item[(3)] $su$ region:
	heavy quarks scattering off thermal gluons via $s$- and $u$-channels for both small and large momentum transfers;
	this contribution is generally small compared to the $t$-channel.
\end{enumerate}
The total scattering rate is then
\begin{equation}\label{eq:Gamma_Total}
\Gamma=\Gamma_{(t)}^{soft}+\Gamma_{(t)}^{hard}+\Gamma_{(su)}.
\end{equation}
The first two terms on the right-hand side are the soft and hard contributions from the $t$-channel, respectively,
and the third term is the contribution from the $s$- and $u$-channels.

\subsection{Collisional energy loss and rransport coefficients of heavy quarks}
The energy loss per unit path length for a heavy quark is given by
\begin{equation}\label{eq:ELoss_Def}
-\frac{dE}{dz}=\int d^3\vec{q}\frac{d\Gamma}{d^3\vec{q}}\frac{\omega}{v},
\end{equation}
where $v$ is the velocity of the heavy quark and $d\Gamma/d^3\vec{q}$
is the differential scattering rate with respect to the three-momentum transfer $\vec{q}$ and energy transfer $\omega$.
Within the SHFM, the total energy loss is obtained by substituting Eq.~(\ref{eq:Gamma_Total}) into Eq.~(\ref{eq:ELoss_Def}),
\begin{equation}\label{eq:ELoss_Sum}
-\frac{dE}{dz}=\left[-\frac{dE}{dz}\right]_{(t)}^{soft}+\left[-\frac{dE}{dz}\right]_{(t)}^{hard}+\left[-\frac{dE}{dz}\right]_{(su)}.
\end{equation}

The interactions between heavy quarks (HQs) and the medium partons
are also encoded into the momentum diffusion coefficients:
\begin{equation}
\begin{aligned}\label{eq:KappaT_Def}
\kappa_{T} &= \frac{1}{2}  \int d^{3}\vec{q} \; \frac{d\Gamma}{d^{3}\vec{q}} \; \vec{q}^{\;2}_{T}= \frac{1}{2}  \int d^{3}\vec{q} \; \frac{d\Gamma}{d^{3}\vec{q}}\biggr[ \omega^{2} - t - \frac{(2\omega E_{1}-t)^{2}}{4\vec{p}_{1}^{\;2}} \biggr],
\end{aligned}
\end{equation}
\begin{equation}
\begin{aligned}\label{eq:KappaL_Def}
\kappa_{L} &= \int d^{3}\vec{q} \frac{d\Gamma}{d^{3}\vec{q}} \; q^{2}_{L}
= \frac{1}{4\vec{p}_{1}^{\;2}} \int d^{3}\vec{q} \frac{d\Gamma}{d^{3}\vec{q}} \; (2\omega E_{1}-t)^{2},
\end{aligned}
\end{equation}
which quantifies the momentum fluctuations perpendicular (transverse)
and parallel (longitudinal) to the direction of propagation, respectively.
Within the Langevin transport framework~\cite{CTGUHybrid2,CTGUHybrid5},
the drag coefficient $\eta_D$—related to energy loss—can be derived from $\kappa_T$ and $\kappa_L$ via
the dissipation-fluctuation relation $\eta_D = \eta_D(\kappa_T, \kappa_L)$;
see Ref.~\cite{Li:2019lex} for details.
Analogous to Eqs.~(\ref{eq:Gamma_Total}) and (\ref{eq:ELoss_Sum}),
the full expressions for these coefficients are derived from the scattering rate.

As can be seen from Eqs.~(\ref{eq:ELoss_Def}), (\ref{eq:KappaT_Def}), and (\ref{eq:KappaL_Def}),
the scattering rate [Eq.~(\ref{eq:Gamma_Total})] is the central quantity in this work.
In the following, we compute the scattering rate using the soft-hard decomposition model,
incorporating a thermal perturbative description for soft scattering ($-t < -t^\ast$)
and a pQCD description for hard scattering ($-t > -t^\ast$).
For completeness, we present key results with necessary explanations;
detailed derivations can be found in Refs.~\cite{li2021langevin,peng2024unraveling}.

\subsubsection{Soft region ($-t < -t^{\ast}$)}\label{subsubsec:SoftReg}
By extending Weldon's model~\cite{Weldon:1983jn} to finite-$\mu$,
the scattering rate for soft scattering between a heavy quark and thermal partons
can be obtained from the imaginary part of the heavy quark self-energy:
\begin{equation}\label{eq:Weldon}
\Gamma(E_1,T,\mu)=-\frac{1}{2E_1}\bigr[1-n_{F}^{+}(E_1,T,\mu)\bigr] Tr\bigr[(\slashed{P}_1+m_1) \cdot Im\Sigma(E_1+i\epsilon,\vec{p}_1)\bigr],
\end{equation}
where $m_1$ is the mass of the incoming heavy quark.
The self-energy in Minkowski space is given by
\begin{equation}
\begin{aligned}\label{eq:Sigma_Soft1}
{\Sigma}(P_{1}) &= iC_{F}g^{2} \int\frac{d^{4}Q}{(2\pi)^{4}} G^{\mu\nu}(Q) \gamma_{\mu} S(K) \gamma_{\nu} \\
&= -C_{F}g^{2}T \sum_{q^{0}}^{\;} \int\frac{d^{3}\vec{q}}{(2\pi)^{3}} G^{\mu\nu}(Q) \gamma_{\mu} S(K) \gamma_{\nu}.
\end{aligned}
\end{equation}
Here, $C_{F}=(N_{c}^{2}-1)/(2N_{c})$ is the quark Casimir factor for the fundamental representation,
$g=\sqrt{4\pi\alpha_s}$ is the strong coupling constant,
and $Q=(q^{0},\vec{q}\;)$ is the four-momentum transfer.
In the second equality of Eq.~(\ref{eq:Sigma_Soft1}),
the integral over $q^0$ is replaced by a sum over bosonic Matsubara frequencies
$q^0 = i\omega_n = i2n\pi T$ ($n \in \mathbb{Z}$).
The HTL-resummed gluon propagator in Coulomb gauge is~\cite{JeanPR02}
\begin{equation}\label{eq:GluonPrp_HTL}
G^{\mu\nu}(Q)=-G_{L}(Q)\delta^{\mu0}\delta^{\nu0}-G_{T}(Q)\left(\delta^{ij}-\widehat{q}^{i}\widehat{q}^{j}\right)\delta^{\mu i}\delta^{\nu j},
\end{equation}
with the transverse and longitudinal components given by inverting
\begin{subequations}
\begin{align}
G_{T}^{-1}(q^{0},\vec{q}\;) &= q_{0}^{2}-\vec{q}^{\;2} - \Pi_{T}, \label{eq:Propagator_Soft_T} \\
G_{L}^{-1}(q^{0},\vec{q}\;) &= \vec{q}^{\;2} + \Pi_{L}. \label{eq:Propagator_Soft_L}
\end{align}
\end{subequations}
The transverse and longitudinal soft gluon self-energies are
\begin{equation}
\Pi_L(q^{0},q)=M_D^2\biggr[1-\frac{q_{0}}{2q}\ln\left(\frac{q_{0}+q}{q_{0}-q}\right)\biggr],
\end{equation}
\begin{equation}
\Pi_T(q^{0},q)=\frac{M_D^2}{2}\left[\frac{q_{0}^2}{q^2}+\left(1-\frac{q_{0}^2}{q^2}\right)\frac{q_{0}}{2q}\ln\left(\frac{q_{0}+q}{q_{0}-q}\right)\right],
\end{equation}
where $q=|\vec{q}\;|$.
The Debye screening mass squared for gluons at finite chemical potential $\mu\ne0$ is~\cite{PhysRevC.71.064901, PhysRevC.54.2588}
\begin{equation}
\begin{aligned}\label{eq:MD_WithMu}
M_D^2(T,\mu) &= -\pi \alpha_{s} d_{g} \int\frac{d^{3}\vec{p}}{(2\pi)^{3}} \frac{\partial}{\partial|\vec{p}\;|} (N_{c}\mathcal{N}_{B} + N_{f}\mathcal{N}_{F})
= m_D^2+\frac{N_fg^2\mu^2}{2\pi^2},
\end{aligned}
\end{equation}
where $m_D^2$ is the Debye mass at vanishing chemical potential $\mu=0$,
\begin{equation}\label{eq:MD_WithOutMu}
m_D^2(T) = \biggr(\frac{N_c}{3}+\frac{N_f}{6}\biggr)g^2 T^2,
\end{equation}
which scales approximately linearly with $T$.
The second term on the right-hand side of Eq.~(\ref{eq:MD_WithMu}) represents the finite-$\mu$ correction.
Here, $N_c = 3$ is the number of colors for $SU(N_{c})$ symmetry,
$d_g = 2(N_c^2 - 1) = 16$ is the gluon degeneracy factor,
and $N_f = 3$ is the number of quark flavors.
Assuming local thermal equilibrium, the distribution functions for bosons (gluons) and fermions (light quarks and antiquarks) are 
\begin{equation}\label{eq:ThermalDis_Boson}
\mathcal{N}_B(E,T) = n_{B}(E,T) = \frac{1}{e^{E/T}-1},
\end{equation}
\begin{equation}\label{eq:ThermalDis_FermionAvg}
\begin{aligned}
&\mathcal{N}_{F}(E,T,\mu) = \frac{n_{F}^{+}+n_{F}^{-}}{2},
\end{aligned}
\end{equation}
with
\begin{equation}\label{eq:ThermalDis_FermionInd}
n_F^{\pm}(E,T,\mu) = \frac{1}{e^{(E\mp\mu)/T} + 1}.
\end{equation}
Note that, as shown in Eq.~(\ref{eq:MD_WithMu}), $\partial \mathcal{N}/\partial |\vec{p}\;|\approx -\mathcal{N}/|\vec{p}\;|$
holds in local equilibrium~\cite{PhysRevC.71.064901, PhysRevC.54.2588}.

Substituting the mixed representations of the bare quark and HTL-resummed gluon propagators
into the soft interaction rate, we arrive at
\begin{align}
\Gamma_{(t)}^{soft}(E_1,T,\mu) &= C_F g^2 \int \frac{d^3\vec{q}}{(2\pi)^3} \int d\omega
[1+n_{B}(\omega,T)] \delta(\omega - \vec{v}_1 \cdot \vec{q})
\left\{ \rho_L + \vec{v}^2 \left[1 - (\widehat{v}\cdot\widehat{q})^2\right] \rho_T \right\}.
\end{align}
The energy loss is
\begin{equation}\label{eq:ELoss_Soft_vsX}
\biggr[-\frac{dE}{dz}\biggr]_{(t)}^{soft}(E_1,T,\mu) = \frac{C_Fg^2}{8\pi^2v_{1}^{2}}\int_{t^{\ast}}^0dt\left(-t\right)\int_0^{v}dx\frac{x}{(1-x^2)^2}\left[\rho_L(t,x,\mu)+(v^2-x^2)\rho_T(t,x,\mu)\right],
\end{equation}
where $\rho_{T/L}(t,x,\mu)$ [Eqs.~(\ref{eq:SpecFunc_T_vsX}) and (\ref{eq:SpecFunc_L_vsX})]
indicates the spectral functions and $x=\omega/q$.
Here, we simply summarize the final results,
and the detailed aspects are relegated to the Appendix~\ref{appendix:Derive_ELoss_Soft_vsX}.

Similarly, the transverse and longitudinal momentum diffusion coefficients are
\begin{equation}
\begin{aligned}\label{eq:kappaT_Soft_vsX}
\biggr[\kappa_T\biggr]^{soft}_{(t)}(E_1,T,\mu) &= \frac{C_F g^2}{16\pi^{2} v_{1}^{3}} \int_{t^{\ast}}^0 dt
(-t)^{3/2} \int_0^v dx  \frac{v^2 - x^2}{(1 - x^2)^{5/2}}
[ \rho_L + (v^2 - x^2) \rho_T]
\coth\left( \frac{x}{2T} \sqrt{\frac{-t}{1 - x^2}} \right),
\end{aligned}
\end{equation}
\begin{equation}
\begin{aligned}\label{eq:kappaL_Soft_vsX}
\biggr[\kappa_{L}\biggr]^{soft}_{(t)}(E_1,T,\mu) &= \frac{C_{F}g^{2}}{8\pi^{2}v_{1}^{3}}\int_{t^{*}}^{0}dt(-t)^{3/2}
\int_{0}^{v}dx\frac{x^{2}}{(1-x^{2})^{5/2}} [\rho_L+(v^2-x^2)\rho_T]
\coth\left(\frac{x}{2T}\sqrt{\frac{-t}{1-x^{2}}}\right),
\end{aligned}
\end{equation}

\subsubsection{Hard region ($-t > -t^{\ast}$)}\label{subsubsec:HardReg}
The transition rate for two-body scattering is defined as the rate
for collisions between the injected heavy quark and a target medium parton $i$,
with momentum transfer $\vec{q}$:
\begin{equation}
\omega_{Qi}(\vec{p}_1,\vec{q}\;)=\int\frac{d^{3}\vec{p}_2}{(2\pi)^{3}}\mathcal{N}_i(E_2,T) \cdot v_{rel}d\sigma_{Qi}(\vec{p}_1,\vec{p}_2\to\vec{p}_3,\vec{p}_4).
\end{equation}
The thermal distributions $\mathcal{N}_i(E_2, T)$ for bosons and fermions
are given by Eqs.~(\ref{eq:ThermalDis_Boson}) and (\ref{eq:ThermalDis_FermionAvg}), respectively;
statistical effects for outgoing particles are neglected.
The spin- and color-averaged differential cross section is
\begin{equation}
\begin{aligned}\label{eq:diff_cross_section}
v_{rel} & d\sigma_{Qi}(\vec{p}_1,\vec{p}_2 \rightarrow \vec{p}_3,\vec{p}_4)
= \frac{1}{2E_1}\frac{1}{2E_2}\frac{d^3\vec{p}_3}{(2\pi)^32E_3}\frac{d^3\vec{p}_4}{(2\pi)^32E_4}\overline{|\mathcal{M}^{2}|}_{Qi}
(2\pi)^4 \delta^{(4)}(P_1+P_2-P_3-P_4),
\end{aligned}
\end{equation}
where $v_{rel}=\sqrt{(P_{1}\cdot P_{2})^{2}-(m_{1}m_{2})^{2}}/(E_{1}E_{2})$ is the relative velocity.
The relevant interaction rate is
\begin{equation}
\begin{aligned}\label{eq:Gamma_Hard}
\Gamma_{Qi(t)}^{hard}(E_{1},T,\mu) &= \int d^{3}\vec{q} ~\omega_{Qi}(\vec{p}_1,\vec{q}\;) \\
&=\frac{1}{2E_{1}}\int\frac{d^{3}\vec{p}_2}{(2\pi)^{3}}
\frac{\mathcal{N}_i(E_2,T)}{2E_{2}}\int\frac{d^{3}\vec{p}_3}{(2\pi)^{3}}\frac{1}{2E_{3}}\int\frac{d^{3}\vec{p}_4}{(2\pi)^{3}}\frac{1}{2E_{4}} 
\overline{|\mathcal{M}^{2}|}_{Qi(t)}(2\pi)^{4}\delta^{(4)}(P_1+P_2-P_3-P_4).
\end{aligned}
\end{equation}
Substituting into Eqs.~(\ref{eq:ELoss_Def}), (\ref{eq:KappaT_Def}),
and (\ref{eq:KappaL_Def}) gives the $t$-channel energy loss and diffusion coefficients:
\begin{equation}
\begin{aligned}\label{eq:ELoss_Hard}
\biggr[-\frac{dE}{dz}\biggr]_{(t)}^{hard}(E_1,T,\mu) &= \sum_{i=q,g}\biggr[-\frac{dE}{dz}\biggr]_{Qi(t)}^{hard} \\
&= \frac{1}{256\pi^{3}p_{1}^{2}}\int_{p_{2,min}}^{\infty}dp_{2}E_{2}\mathcal{N}_i(E_2,T) 
\int_{-1}^{\cos\psi|_{max}}d(\cos\psi)\int_{t_{min}}^{t^{*}}dt\frac{b}{a^{3}} \; \overline{|\mathcal{M}^{2}|}_{Qi(t)},
\end{aligned}
\end{equation}
\begin{equation}                                                                                                                                              
\begin{aligned}\label{eq:KappaT_Hard_t}
\biggr[\kappa_{T}\biggr]^{hard}_{(t)}(E_1,T,\mu) =& \sum_{i=q,g} \biggr[\kappa_{T}\biggr]^{hard}_{Qi(t)} \\
=& \frac{1}{256\pi^{3}p_{1}^{3}E_{1}} \sum_{i=q,g} \int_{p_{2,min}}^{\infty}dp_{2} E_{2} \mathcal{N}_{i}(E_{2},T)
\int_{-1}^{cos\psi|_{max}} d(cos\psi) \\
&\int_{t_{min}}^{t^{\ast}}dt ~\frac{1}{a} \biggr[ -\frac{m_{1}^{2}(D+2b^{2})}{8a^{4}} + \frac{E_{1}tb}{2a^{2}} - t(p_{1}^{2}+\frac{t}{4}) \biggr] \; \overline{|\mathcal{M}^{2}|}_{Qi(t)},
\end{aligned}
\end{equation}
\begin{equation}
\begin{aligned}\label{eq:KappaL_Hard_t}
\biggr[\kappa_{L}\biggr]^{hard}_{(t)}(E_1,T,\mu) =& \sum_{i=q,g} \biggr[\kappa_{L}\biggr]^{hard}_{Qi(t)} \\
=& \frac{1}{256\pi^{3}p_{1}^{3}E_{1}} \sum_{i=q,g} \int_{p_{2,min}}^{\infty}dp_{2} E_{2} \mathcal{N}_{i}(E_{2},T)
\int_{-1}^{cos\psi|_{max}} d(cos\psi) \\
&\int_{t_{min}}^{t^{\ast}} dt ~\frac{1}{a} \biggr[ \frac{E_{1}^{2}(D+2b^{2})}{4a^{4}} -\frac{E_{1}tb}{a^{2}} + \frac{t^{2}}{2} \biggr] \; \overline{|\mathcal{M}^{2}|}_{Qi(t)}.
\end{aligned}
\end{equation}

The vacuum matrix elements for heavy quark scattering off
quarks ($i = q$) and gluons ($i = g$) are~\cite{li2021langevin,Combridge79}
\begin{equation}
\begin{aligned}\label{eq:App_MatrixHQq}
&\overline{|\mathcal{M}^{2}|}_{Qq(t)} =
\frac{16}{9} N_{f} N_{c} g^{4} \biggr[ \frac{2\tilde{s}^{\;2}}{t^{2}}+\frac{2(\tilde{s}+m_{1}^{2})}{t}+1 \biggr],
\end{aligned}
\end{equation}
\begin{equation}
\begin{aligned}\label{eq:App_MatrixHQg_t}
\overline{|\mathcal{M}^{2}|}_{Qg(t)} =& 2(N_{c}^{2}-1) g^{4} \biggr[ -\frac{2\tilde{s}\tilde{u}}{t^{2}} +\frac{m_{1}^{2}(\tilde{s}-\tilde{u})-\tilde{s}\tilde{u}}{t\tilde{s}}
- \frac{m_{1}^{2}(\tilde{s}-\tilde{u})+\tilde{s}\tilde{u}}{t\tilde{u}} \biggr],
\end{aligned}
\end{equation}
\begin{equation}
\begin{aligned}\label{eq:App_MatrixHQg_su}
\overline{|\mathcal{M}^{2}|}_{Qg(su)} =& \frac{8}{9} (N_{c}^{2}-1) g^{4} \biggr[ \frac{2m_{1}^{2}(\tilde{s}+2m_{1}^{2})-\tilde{s}\tilde{u}}{\tilde{s}^{\;2}} +
\frac{2m_{1}^{2}(\tilde{u}+2m_{1}^{2})-\tilde{s}\tilde{u}}{\tilde{u}^{\;2}} - \frac{m_{1}^{2}(4m_{1}^{2}-t)}{4\tilde{s}\tilde{u}} \biggr].
\end{aligned}
\end{equation}
Here, we have introduced the abbreviation,
\begin{equation}
\begin{aligned}
\tilde{s}\equiv s-m_{1}^{2}, \qquad \tilde{u}\equiv u-m_{1}^{2}.
\end{aligned}
\end{equation}
The integration boundaries and auxiliary variables in
Eqs.~(\ref{eq:ELoss_Hard})-(\ref{eq:KappaL_Hard_t}) are:
\begin{subequations}
\begin{align}
&p_{2,min} = \frac{ |t^{\ast}|+\sqrt{(t^{\ast})^{2} + 4m_{1}^{2} |t^{\ast}|} }{4(E_{1}+p_{1})} \label{eq:App_P2Min}, \\
&cos\psi|_{max} = min \biggr\{ 1,~\frac{E_{1}}{p_{1}} - \frac{|t^{\ast}| + \sqrt{(t^{\ast})^{2} + 4m_{1}^{2} |t^{\ast}|}}{4p_{1}p_{2}} \biggr\} \label{eq:App_PhiMax}, \\
&t_{min} = -\frac{(s-m_{1}^{2})^{2}}{s} \label{eq:App_tMin}, \\
&a = \frac{s-m_{1}^{2}}{p_{1}} \label{eq:App_aVal}, \\
&b =-\frac{2t}{p_{1}^{2}} \bigr[ E_{1}(s-m_{1}^{2})-E_{2}(s+m_{1}^{2}) \bigr] \label{eq:App_bVal}, \\
&c =-\frac{t}{p_{1}^{2}} \biggr\{ t\bigr[ (E_{1}+E_{2})^{2}-s \bigr] + 4p_{1}^{2}p_{2}^{2}sin^{2}\psi \biggr\} \label{eq:App_cVal}, \\
&D = b^{2}+4a^{2}c=-t \biggr[ ts + (s-m_{1}^{2})^{2} \biggr] \cdot \biggr( \frac{4E_{2}sin\psi}{p_{1}} \biggr)^{2} \label{eq:App_DVal}.
\end{align}
\end{subequations}

As noted in Sec.~\ref{sec:gamma_HTLpQCD},
due to the finite heavy quark masses ($m_c = 1.5~\text{GeV}$, $m_b = 4.75~\text{GeV}$),
the $s$- and $u$-channel cross sections for heavy quark-gluon scattering are not divergent at small momentum transfer.
Thus, the energy loss and momentum diffusion coefficients for these channels can be obtained
by modifying the $t$-channel integrals [Eqs.~(\ref{eq:ELoss_Hard})-(\ref{eq:KappaL_Hard_t})] as follows:
(i) set $|\vec{p}_2|_{\text{min}} = 0$ in Eq.~(\ref{eq:App_P2Min});
(ii) set $\cos\psi|_{\text{max}} = 1$ in Eq.~(\ref{eq:App_PhiMax});
(iii) set $t^\ast = 0$ in Eq.~(\ref{eq:App_tMin}).
This yields
\begin{equation}
\begin{aligned}\label{eq:dEdz_Hard_su_Perturbative}
\biggr[-\frac{dE}{dz}\biggr]^{hard}_{(su)}(E_1,T) &= \frac{1}{256\pi^{3}p_{1}^{2}} \int_{0}^{\infty}dp_{2} E_{2} n_{B}(E_{2},T)
\int_{-1}^{1} d(cos\psi) \int_{t_{min}}^{0}dt \frac{b}{a^{3}} \; \overline{|\mathcal{M}^{2}|}_{(su)},
\end{aligned}
\end{equation}
\begin{equation}
\begin{aligned}\label{eq:KappaT_Hard_su_Perturbative}
\biggr[\kappa_{T}\biggr]^{hard}_{(su)}(E_1,T) =& \frac{1}{256\pi^{3}p_{1}^{3}E_{1}}
\int_{0}^{\infty}dp_{2} E_{2} n_{B}(E_{2},T) \int_{-1}^{1} d(cos\psi) \\
& \int_{t_{min}}^{0}dt \frac{1}{a} \biggr[ -\frac{m_{1}^{2}(D+2b^{2})}{8a^{4}} + \frac{E_{1}tb}{2a^{2}} - t(p_{1}^{2}+\frac{t}{4}) \biggr] \; \overline{|\mathcal{M}^{2}|}_{(su)},
\end{aligned}
\end{equation}
\begin{equation}
\begin{aligned}\label{eq:KappaL_Hard_su_Perturbative}
\biggr[\kappa_{L}\biggr]^{hard}_{(su)}(E_1,T) =& \frac{1}{256\pi^{3}p_{1}^{3}E_{1}}
\int_{0}^{\infty}dp_{2} E_{2} n_{B}(E_{2},T) \int_{-1}^{1} d(cos\psi) \\
& \int_{t_{min}}^{0} dt \frac{1}{a} \biggr[ \frac{E_{1}^{2}(D+2b^{2})}{4a^{4}}
- \frac{E_{1}tb}{a^{2}} + \frac{t^{2}}{2} \biggr] \; \overline{|\mathcal{M}^{2}|}_{(su)}.
\end{aligned}
\end{equation}

\section{Energy loss in the High-Energy Approximation (HEA)}\label{sec:HEA}
We note that the integrals in Eqs.~(\ref{eq:ELoss_Soft_vsX}), (\ref{eq:ELoss_Hard})                                                                           
and (\ref{eq:dEdz_Hard_su_Perturbative}) are difficult to evaluate analytically
for arbitrary heavy-quark energy $E_{1}$, medium temperature $T$ and chemical potential $\mu$.
The physical interpretations are thus challenging in general.
Building on our previous work~\cite{peng2024unraveling},
we derive analytic expressions for the energy loss at finite-$\mu$ in the high-energy limit ($E_1 \gg m_1^2 / T$).
In this regime, the heavy quark velocity $v_1 = |\vec{p}_1| / E_1 \to 1$ since $E_1 \approx p_1 \gg m_1$.
The target partons have energies $E_2 \sim \mathcal{O}(T)$ because they are in thermal equilibrium at a temperature $T$.
The Mandelstam variable $s$ satisfies 
\begin{equation}
        \begin{aligned}\label{eq:App_sVar_HEA}
                &s \sim \mathcal{O}(E_{1}T) \gg m^{2}_{1} \gg -t^{\ast}, \\
                &\tilde{s}\equiv s-m^{2}_{1} \approx s.
        \end{aligned}
\end{equation}
For $t$ and $u$, we have
\begin{equation}
        \begin{aligned}
                &t_{min}=-\tilde{s}+\frac{m^{2}_{1}\tilde{s}}{s} \approx -\tilde{s}, \\
                &m^{2}_{1} < -\tilde{u} < \tilde{s}.
        \end{aligned}
\end{equation}
The parameters in Eqs.~(\ref{eq:App_P2Min})-(\ref{eq:App_bVal}) simplify to
\begin{equation}
        \begin{aligned}\label{eq:App_MidPara_HEA}
                &|\vec{p}_{2}|_{min} \approx 0, \qquad cos\psi|_{max} \approx 1, \qquad t_{min} \approx -s,
                \qquad a \approx \frac{s}{E_{1}}, \qquad b \approx -\frac{2ts}{E_{1}}.
        \end{aligned}
\end{equation}
Further details can be found in Ref.~\cite{peng2024unraveling}.
We use these approximations to simplify the energy loss expressions
in Eqs.~(\ref{eq:ELoss_Soft_vsX}) and (\ref{eq:ELoss_Hard}).

In soft collisions, the energy loss becomes
\begin{equation}
\begin{aligned}\label{eq:App_dEdz_Tmp1}
&\biggr[-\frac{dE}{dz}\biggr]^{soft}_{(t)}(E_{1},T,\mu) \approx \frac{C_{F}g^{2}}{8\pi^{2}} \int_{0}^{1}dx \int^{0}_{t^{*}} dt \; (-t)
\biggr\{ \frac{x}{(1-x^{2})^{2}} \bigr[ \rho_{L}(t,x,\mu) + (1-x^{2})\rho_{T}(t,x,\mu) \bigr] \biggr\}.
\end{aligned}
\end{equation}
In the weak-coupling limit $M_D^2 \ll -t^\ast \ll T^2$,
the integrals over $t$ and $x$ can be further simplified (see Appendix B of Ref.~\cite{peng2024unraveling}),
yielding
\begin{equation}\label{eq:ELoss_Soft_HEA}
\biggr[-\frac{dE}{dz}\biggr]_{(t)}^{soft-HEA}(E_{1},T,\mu) \approx \frac{C_F}{16\pi}g^2m_D^2\ln\frac{-2t^{\ast}}{m_D^2} + \mathcal{F}_{1},
\end{equation}
where
\begin{equation}
\mathcal{F}_{1}(T,\mu) = \frac{C_{F}g^{2}}{16\pi} \biggr( \frac{N_{f}g^{2}\mu^{2}}{2\pi^{2}}
\ln\frac{-2t^{*}}{M_{D}^{2}}-m_{D}^{2}\ln\frac{M_{D}^{2}}{m_{D}^{2}} \biggr).
\end{equation}
The first term on the right-hand side of Eq.~(\ref{eq:ELoss_Soft_HEA}) is the result for $\mu = 0$~\cite{peng2024unraveling},
and the second term is the finite-$\mu$ correction;
note that $\mathcal{F}_1(\mu = 0) = 0$.

In hard collisions, the energy loss for $Qi$ ($i = q, g$) scattering in the $t$-channel becomes
\begin{equation}\label{eq:ELoss_Hard_tCh_Def}
\biggr[-\frac{dE}{dz}\biggr]_{Qi(t)}^{hard-HEA}(E_{1},T,\mu) \approx \int\frac{d^3\vec{p}_2}{(2\pi)^3}\frac{\mathcal{N}_i(E_2,T)}{2E_2}
\int_{-t^{\ast}}^s d(-t)(-t) \frac{\overline{|\mathcal{M}^{2}}|_{Qi(t)}^{HEA}}{16\pi\tilde{s}^2}.
\end{equation}
The matrix elements in Eqs.~(\ref{eq:App_MatrixHQq}) and (\ref{eq:App_MatrixHQg_t}) simplify to
\begin{equation}
\begin{aligned}\label{eq:App_MatrixHQq_HEA}
\overline{|\mathcal{M}^2|}_{Qq(t)}^{HEA} \approx& \frac{16}{9} N_{f} N_{c} g^{4} \biggr( \frac{2\tilde{s}^{\;2}}{t^{2}}+\frac{2\tilde{s}}{t}+1 \biggr),
\end{aligned}
\end{equation}
\begin{equation}
\begin{aligned}\label{eq:App_MatrixHQg_t_HEA}
\overline{|\mathcal{M}^{2}|}_{Qg(t)}^{HEA} \approx& 2(N_{c}^{2}-1) g^{4} \biggr( \frac{2\tilde{s}^{\;2}}{t^{2}}+\frac{2\tilde{s}}{t}+1 \biggr).
\end{aligned}
\end{equation}
Substituting Eq.~(\ref{eq:App_MatrixHQq_HEA}) into Eq.~(\ref{eq:ELoss_Hard_tCh_Def}) and using
\begin{equation}
\begin{aligned}\label{eq:App_InteFunc_Fermi1}
\int_0^\infty dE\;E\mathcal{N}_F(E,T,\mu)=\frac{\pi^2 T^2}{12}+\frac{\mu^2}{4},
\end{aligned}
\end{equation}
\begin{equation}
\begin{aligned}\label{eq:App_InteFunc_Fermi2}
\int_{0}^{\infty}dE\;E\mathcal{N}_{F}(E,T,\mu)\ln\frac{E}{\mathcal{A}} = \frac{\pi^{2}T^{2}}{12}\biggr[\ln\frac{2T}{\mathcal{A}}+1-\gamma_{E}+\frac{\zeta^{\prime}(2)}{\zeta(2)}\biggr] + \mathcal{F}_{2},
\end{aligned}
\end{equation}
with
\begin{equation}
\begin{aligned}\label{eq:Midd_F2}
\mathcal{F}_{2}(T,\mu) =& \frac{\mu^{2}}{4}\left(\ln\frac{T}{\mathcal{A}} + 1 - \gamma_{E}\right)
- \frac{T^{2}}{2} \biggr[\left.\frac{\partial}{\partial s}Li_{s}(-e^{-\mu/T})\right|_{s=2}+\left.\frac{\partial}{\partial s}Li_{s}(-e^{\mu/T})\right|_{s=2}\biggr] - \frac{\pi^{2}T^{2}}{12}\biggr[\ln2+\frac{\zeta^{\prime}(2)}{\zeta(2)}\biggr],
\end{aligned}
\end{equation}
we obtain the energy loss for $Qq$ in $t$-channel scattering:
\begin{equation}
\begin{aligned}\label{eq:ELoss_Hard_Qq_HEA}
\biggr[-\frac{dE}{dz}\biggr]_{Qq(t)}^{hard-HEA}(E_{1},T,\mu) \approx& \frac{N_fN_c }{216\pi}g^4T^2 \biggr[ \ln\frac{8E_1T}{-t^{\ast}}-\frac{3}{4}-\gamma+\frac{\zeta^{\prime}(2)}{\zeta(2)}\biggr] + \mathcal{F}_{3},
\end{aligned}
\end{equation}
where
\begin{equation}
\begin{aligned}\label{eq:Midd_F3}
\mathcal{F}_{3}(E_{1},T,\mu) =& \frac{N_{f}N_{c}}{18\pi^{3}}g^{4} \left[ \mathcal{F}_{2}\big|_{\mathcal{A}=-t^{\ast}/(4E_{1})} - \frac{7\mu^{2}}{16} \right].
\end{aligned}
\end{equation}
Here, $\mathcal{A}$ is independent of $E$ [Eq.~(\ref{eq:App_InteFunc_Fermi2})],
$\gamma_E \approx 0.57722$ is the Euler-Mascheroni constant,
$\zeta$ is the Riemann zeta function with $\zeta'(2)/\zeta(2) \approx -0.56996$,
and $Li_s(z)$ is the polylogarithm function.
The first term on the right-hand side of Eq.~(\ref{eq:ELoss_Hard_Qq_HEA}) is the $\mu = 0$ result~\cite{peng2024unraveling},
and the second term is the finite-$\mu$ correction; $\mathcal{F}_3(\mu = 0) = 0$.

For $Qg$ scattering in the $t$-channel, using
\begin{equation}
\begin{aligned}\label{eq:App_InteFunc_Bose1}
&\int_{0}^{\infty}dE\; E n_{B}(E,T) = \frac{\pi^{2}T^{2}}{6},
\end{aligned}
\end{equation}
\begin{equation}
\begin{aligned}\label{eq:App_InteFunc_Bose2}
&\int_{0}^{\infty}dE \;E n_{B}(E,T) ln\frac{E}{\mathcal{A}} = \frac{\pi^{2}T^{2}}{6}\biggr[ ln\frac{T}{\mathcal{A}}+1-\gamma_{E}+\frac{\zeta^{'}(2)}{\zeta(2)}\biggr],
\end{aligned}
\end{equation}
we find~\cite{peng2024unraveling},
\begin{equation}
\begin{aligned}\label{eq:ELoss_Hard_Qg_t_HEA}
\left[-\frac{dE}{dz}\right]_{Qg(t)}^{hard-HEA}(E_{1},T) \approx& \frac{N_c^2-1}{96\pi}g^4T^2\left[\ln\frac{4E_1T}{-t^{\ast}}-\frac{3}{4}- \gamma_E + \frac{\zeta'(2)}{\zeta(2)}\right].
\end{aligned}
\end{equation}

For $Qg$ scattering in $su$-channel, as shown in Eq.~(\ref{eq:dEdz_Hard_su_Perturbative}),
the energy loss becomes
\begin{equation}\label{eq:ELoss_Hard_suCh_Def}
\biggr[-\frac{dE}{dz}\biggr]_{Qg(su)}^{hard-HEA}(E_{1},T) \approx \int\frac{d^3\vec{p}_2}{(2\pi)^3}\frac{n_B(E_2,T)}{2E_2}
\int_{0}^s d(-t)(-t) \frac{\overline{|\mathcal{M}^{2}}|_{Qg(su)}^{HEA}}{16\pi\tilde{s}^2},
\end{equation}
with the matrix element
\begin{equation}
\begin{aligned}\label{eq:App_MatrixHQg_su_HEA}
\overline{|\mathcal{M}^{2}|}_{Qg(su)}^{HEA}(E_{1},T) \approx& \frac{8}{9} (N_{c}^{2}-1) g^{4} \left( -\frac{\tilde{u}}{\tilde{s}}+\frac{\tilde{s}}{-\tilde{u}} \right).
\end{aligned}
\end{equation}
Following a similar treatment to that for the $t$-channel, we obtain~\cite{peng2024unraveling}
\begin{equation}
\begin{aligned}\label{eq:ELoss_Hard_Qg_su_HEA}
\left[-\frac{dE}{dz}\right]_{Qg(su)}^{hard-HEA}(E_{1},T) \approx& \frac{N_c^2-1}{432\pi}g^4T^2\left[\ln\frac{4E_1T}{m_1^2}-\frac{5}{6}- \gamma_E + \frac{\zeta'(2)}{\zeta(2)}\right],
\end{aligned}
\end{equation}

Combining contributions from the soft [Eq.~(\ref{eq:ELoss_Soft_HEA})],
hard [Eqs.~(\ref{eq:ELoss_Hard_Qq_HEA}) and (\ref{eq:ELoss_Hard_Qg_t_HEA})],
and $su$ [Eq.~(\ref{eq:ELoss_Hard_Qg_su_HEA})] regions,
the total energy loss in the high-energy approximation is
\begin{equation}
\begin{aligned}\label{eq:ELoss_All_HEA}
\left[-\frac{dE}{dz}\right]_{Qq+Qg}^{HEA}(E_{1},T,\mu) \approx& \frac{4}{3}\pi\alpha_{s}^{2}T^{2}\left[\left(1+\frac{N_{f}}{6}\right)\ln\frac{E_{1}T}{m_{D}^{2}}+\frac{2}{9}\ln\frac{E_{1}T}{m_{1}^{2}}+d(N_{f})\right] + \mathcal{G},
\end{aligned}
\end{equation}
where $d(N_{f})\approx 0.146N_{f}+0.05$,
and $\mathcal{G}\left(E_{1},T,\mu\right)$ contains the finite-$\mu$ corrections:
\begin{equation}
\begin{aligned}\label{eq:Midd_G}
\mathcal{G}(E_{1},T,\mu) =& \mathcal{F}_{1}(T,\mu) + \mathcal{F}_{3}(E_{1},T,\mu) \\
\approx& \frac{2\alpha_{s}^{2}}{3\pi}\biggl\{N_{f}\mu^{2}\biggl(\ln\frac{8E_{1}T}{M_{D}^{2}}-\gamma_{E}-\frac{3}{4}\biggr)
-\left.2N_{f}T^{2}\left[\frac{\partial}{\partial s}Li_{s}(-e^{-\mu/T})\right|_{s=2}+\left.\frac{\partial}{\partial s}Li_{s}(-e^{\mu/T})\right|_{s=2}\right] \\
&- \frac{N_{f}\pi^{2}T^{2}}{3}\left[\ln2+\frac{\zeta^{\prime}(2)}{\zeta(2)}\right] - \frac{\pi}{2\alpha_{s}}m_{D}^{2}\ln\frac{M_{D}^{2}}{m_{D}^{2}}\biggr\}.
\end{aligned}
\end{equation}
The first term on the right-hand side of Eq.~(\ref{eq:ELoss_All_HEA})
is the $\mu = 0$ result~\cite{peng2024unraveling},
and the second term is the finite-$\mu$ contribution;
$\mathcal{G}(\mu = 0) = 0$.
The arbitrary scale $t^\ast$ cancels in the full result for both $\mu=0$ and
$\mu\ne0$, as expected~\cite{peng2024unraveling,braaten1991energy,peigne2008collisional,PhysRevD.77.114017}.

\section{Results and discussions}\label{sec:results}
\begin{figure*}[!htbp]
\begin{center}
\setlength{\abovecaptionskip}{-0.1mm}
\includegraphics[width=.48\textwidth]{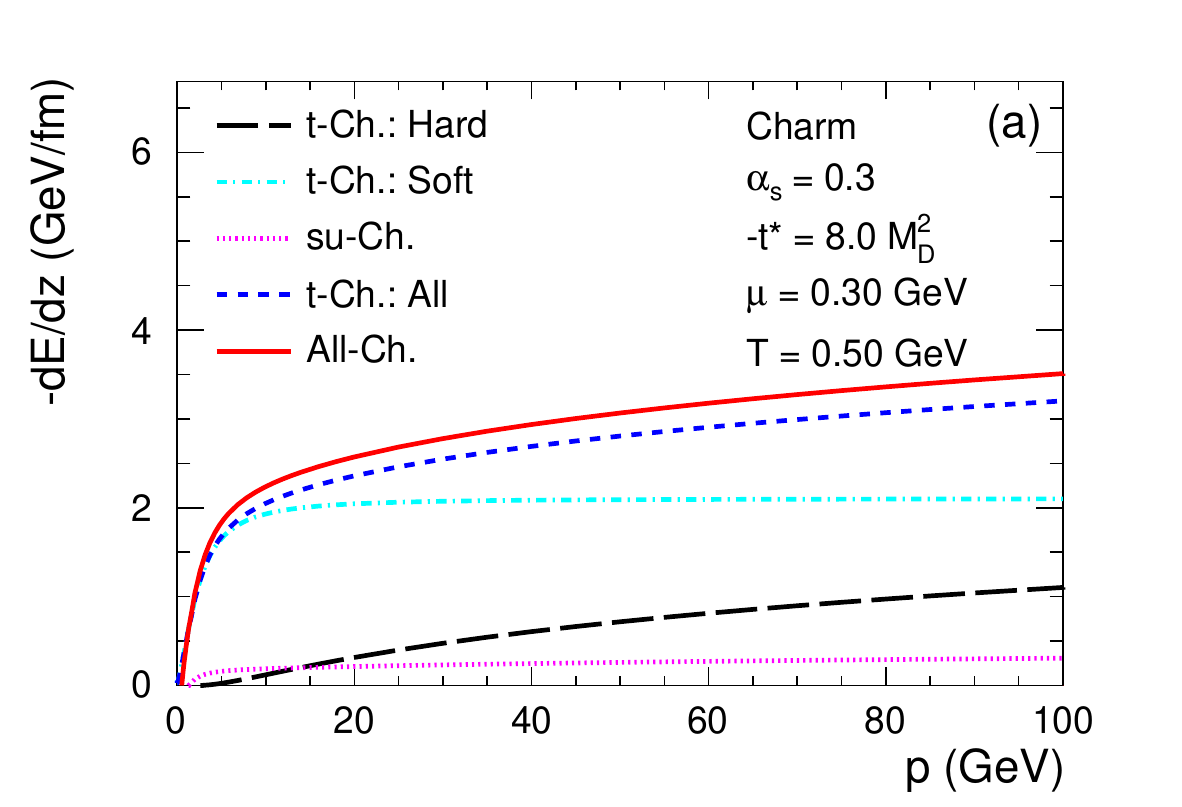}
\includegraphics[width=.48\textwidth]{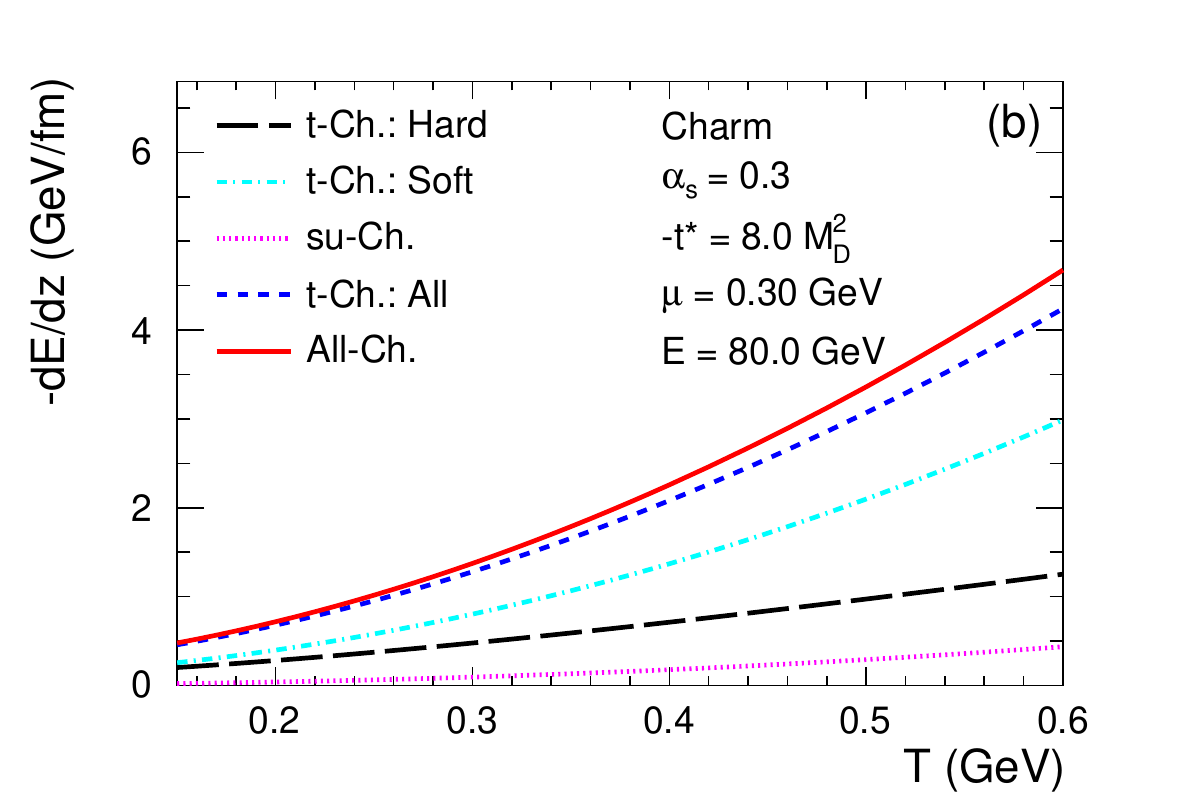}
\caption{(Color online) Left (a): comparison of the energy loss $dE/dz$ as a function of heavy-quark momentum,
for charm quark with $\alpha_{s}=0.3$, $-t^{\ast}=8M_{D}^{2}$ and $\mu=0.3~\rm{GeV}$ at a given temperature $T=0.5~{\rm GeV}$,
contributed by soft interactions in $t$-channel [dot-dashed cyan curve; Eq.~(\ref{eq:ELoss_Soft_vsX})],
and hard interactions in $t$-channel [long-dashed black curve; Eq.~(\ref{eq:ELoss_Hard})],
$su$-channels [dotted pink curve; Eq.~(\ref{eq:dEdz_Hard_su_Perturbative})].
The combined results, i.e. the contributions from the soft and hard interactions in $t$-channels (dashed blue curve)
and from the all components (solid red curve), are shown for comparison.
Right (b): same as panel-a but for $dE/dz$ as a function of temperature at a given heavy-quark energy $E=80~{\rm GeV}$.}
\label{fig:Charm_dEdz_Full_variousChannels}
\end{center}
\end{figure*}

Figure~\ref{fig:Charm_dEdz_Full_variousChannels}(a) presents the collisional energy loss
of a charm quark as a function of its momentum,
computed with the coupling $\alpha_{s}=0.3$, the intermediate cutoff $-t^{\ast}=8M_{D}^{2}$,
the chemical potential $\mu=0.3~\rm{GeV}$ and medium temperature $T=0.5~{\rm GeV}$.
The contributions from different scattering channels are
illustrated by curves with different styles, as detailed in the legend.
We find that all the components show a monotonously rising momentum dependence.
The soft contribution (dot-dashed cyan curve) dominates in the considered momentum region.
It has a stronger energy dependence in the region $p\lesssim 4~{\rm GeV}$,
followed by an almost flat behavior at $p\gtrsim 10~{\rm GeV}$.
This is mainly induced by the fact that,
comparing with the diffusion term,
the drag term dominates the heavy-quark scatterings off thermal partons
since the initial momentum spectra of heavy quarks are much harder than that of medium partons.
Consequently, the energy loss can be described by the drag force,
which is proportional to the heavy-quark velocity $v=p/\sqrt{p^{2}+m^{2}_{\rm Q}}$.
The heavy-quark velocity and energy loss change significantly with increasing momentum at $p\lesssim2-3m_{\rm Q}$,
where the relativistic effect is trivial.
The velocity is close to unity $v\sim1$ at much larger momentum $p\gg m_{\rm Q}$,
where the ultrarelativistic effect should be considered,
and thus the energy loss increases slowly in this regime.

Figure~\ref{fig:Charm_dEdz_Full_variousChannels}(b) shows the temperature dependence of
the charm energy loss at a given heavy-quark energy $E=80~{\rm GeV}$.
The results indicate that all the components show a monotonously rising temperature dependence.
This is because a higher temperature increases the frequency
of momentum-transferring collisions with medium partons,
leading to larger energy loss.
The contributions from $su$-channels (dotted pink curve)
are, as expected, negligible comparing with the results from $t$-channel (dashed blue curve).
Our examination reveals that the above results exhibit low sensitivity
to the choice of the intermediate cutoff $t^{\ast}$.
The same conclusions can be drawn for bottom quarks.

\begin{figure*}[!htbp]
\begin{center}
\setlength{\abovecaptionskip}{-0.1mm}
\includegraphics[width=.48\textwidth]{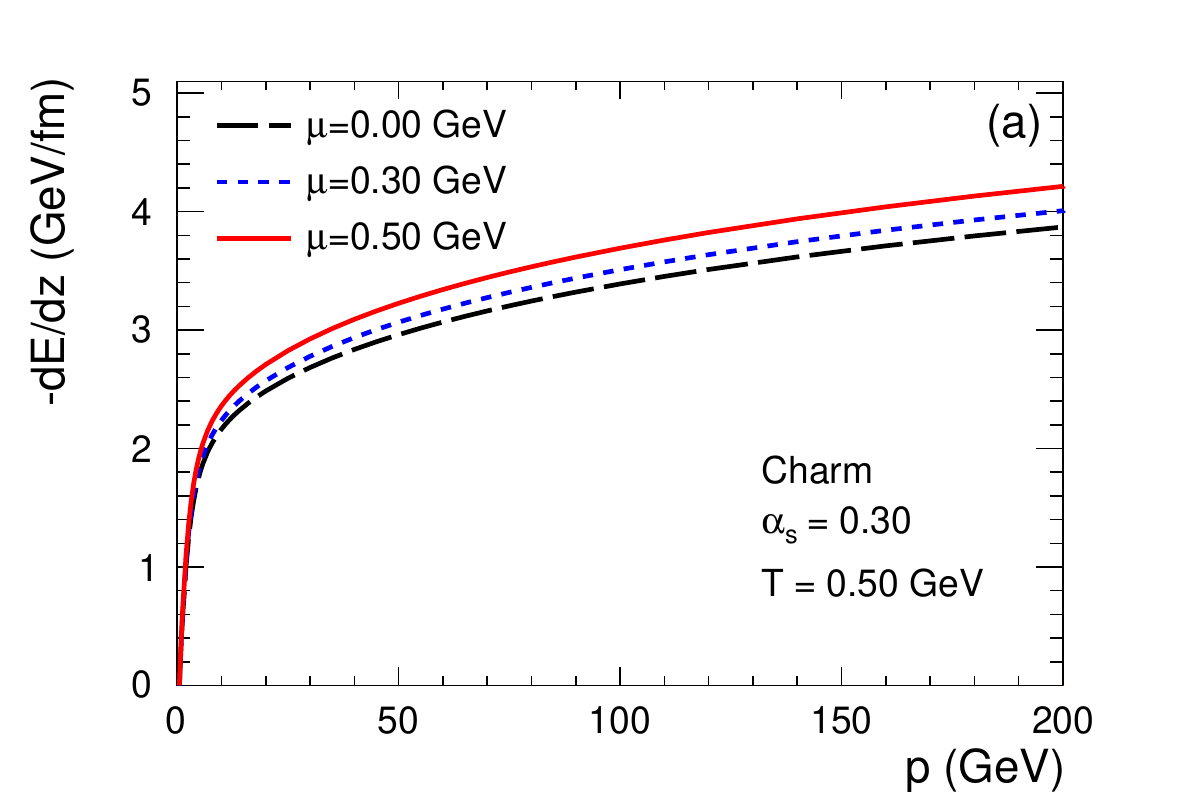}
\includegraphics[width=.48\textwidth]{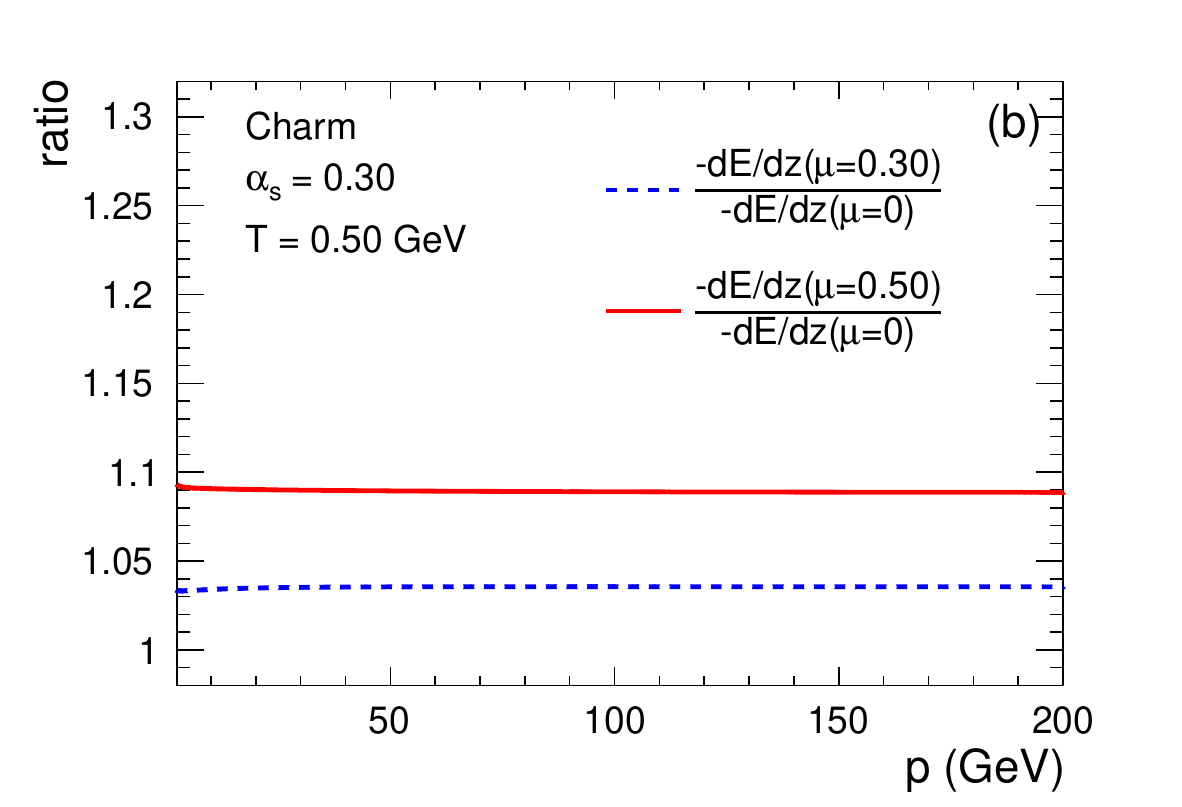}
\caption{(Color online) Left (a): comparison of the total energy loss $dE/dz$ of a charm quark
as a function of its momentum, displaying separately the results with different chemical potential values:
$\mu=0$ (long-dashed black curve), 0.3 GeV (dotted blue curve) and 0.5 GeV (solid red curve).
Right (b): comparison of $dE/dz$ obtained at finite chemical potential ($\mu\ne0$)
with respect to the one at vanishing chemical potential ($\mu=0$).}
\label{fig:Charm_dEdz_AllChan_variousMu_vsP}
\end{center}
\end{figure*}
We now analyze the influence of the chemical potential on the total collisional energy loss.
Figure~\ref{fig:Charm_dEdz_AllChan_variousMu_vsP}(a) displays the momentum dependence of the charm quark $dE/dz$
with $\alpha_{s}=0.3$, $T=0.5~\rm{GeV}$ and various chemical potential values:
$\mu=0$ (long-dashed black curve), $\mu=0.3~\rm{GeV}$ (dotted blue curve) and $\mu=0.5~\rm{GeV}$ (solid red curve).
The ratios of the finite-$\mu$ results to the $\mu=0$ case are shown in Fig.~\ref{fig:Charm_dEdz_AllChan_variousMu_vsP}(b).
When comparing $-dE/dz(\mu\ne0)$ with $-dE/dz(\mu=0)$,
we find that, the chemical potential increases the energy loss.
The enhancement for the result with $\mu=0.5~\rm{GeV}$
reaches $\sim9\%$ at maximum in the desired momentum intervals.
This is because the Debye screening mass increases with chemical potential [Eq.~({\ref{eq:MD_WithMu}})],
and more long-wavelength interactions are therefore screened,
resulting in a larger thermalization rate and energy loss of heavy quarks.
Additionally, as introduced in Sec.~\ref{subsubsec:SoftReg},
the thermal distribution of fermion is modified by the chemical potential $\mu$ [Eq.~(\ref{eq:ThermalDis_FermionAvg})].
For finite $\mu$ it can be expressed as
\begin{equation}\label{eq:ThermalDis_Fermion_Cmp}
\mathcal{N}_{F}(E,T,\mu) = {n}_{F} + \frac{\mu^{2}}{2T^{2}}{n}_{F}(1-{n}_{F})\tanh\frac{E}{2T} + \mathcal{O}(\mu^{4}), \nonumber
\end{equation}
where
\begin{equation}\label{eq:ThermalDis_Fermion_Mu0}
{n}_{F}(E,T) = \frac{1}{e^{E/T}+1} = \mathcal{N}_{F}(E,T,\mu=0), \nonumber
\end{equation}
is the thermal distribution of fermion in the absence of chemical potential $\mu=0$.
We find that the chemical potential increases the fermion density,
resulting in a larger interaction rate [Eq.~(\ref{eq:Gamma_Hard})]
and consequently a stronger energy loss [Eq.~(\ref{eq:ELoss_Hard})].

\begin{figure*}[!htbp]
\begin{center}
\setlength{\abovecaptionskip}{-0.1mm}
\includegraphics[width=.48\textwidth]{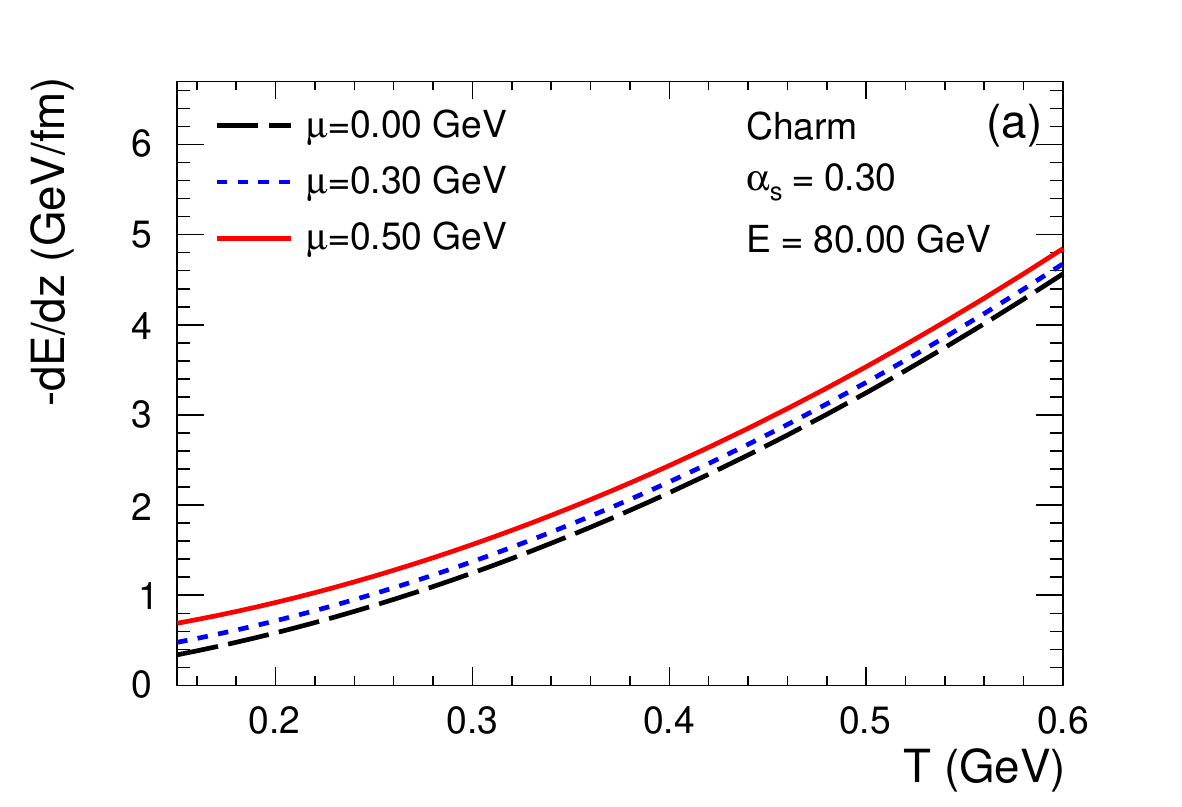}
\includegraphics[width=.48\textwidth]{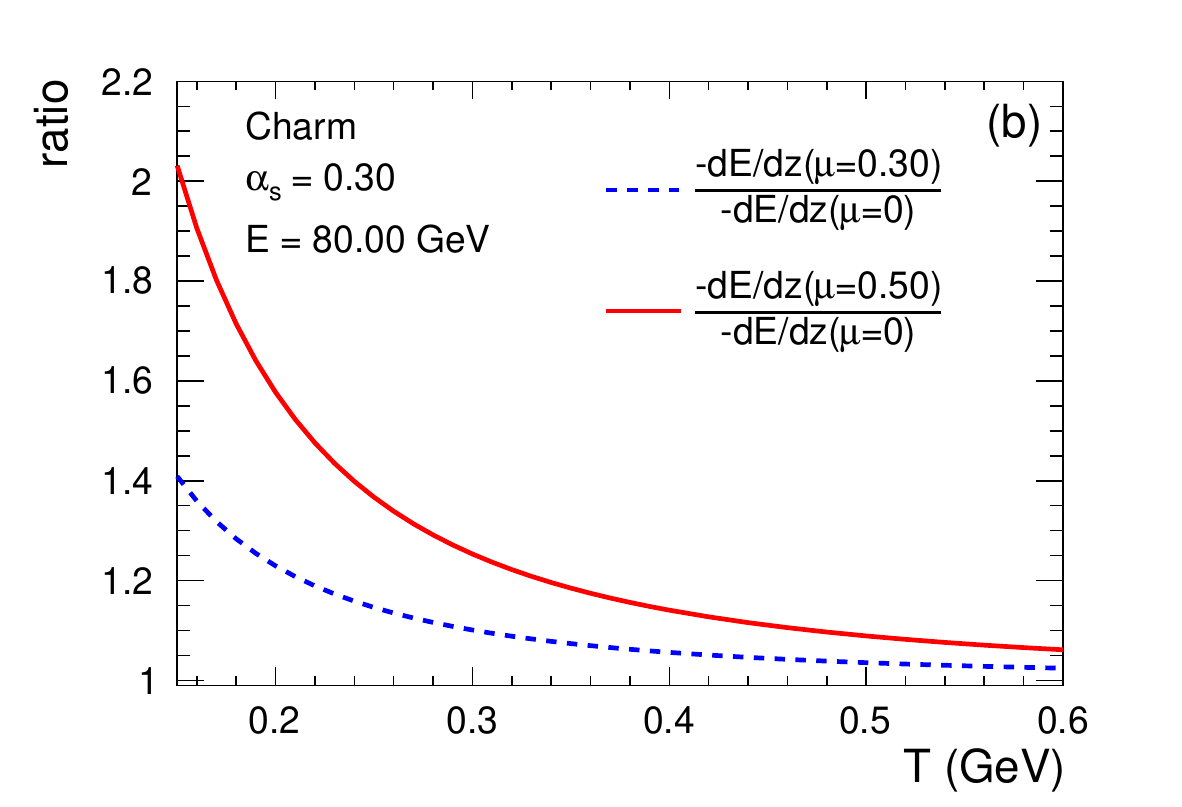}
\caption{Same as Fig.~\ref{fig:Charm_dEdz_AllChan_variousMu_vsP} but as a function of temperature.}
\label{fig:Charm_dEdz_AllChan_variousMu_vsT}
\end{center}                                                                                                                                                  
\end{figure*}
Figure~\ref{fig:Charm_dEdz_AllChan_variousMu_vsT} shows the same results
as in Fig.~\ref{fig:Charm_dEdz_AllChan_variousMu_vsT} but as a function of QCD medium temperature.
The ratios (panel-b in Fig.~\ref{fig:Charm_dEdz_AllChan_variousMu_vsT}) present
a strong (weak) $T$-dependence at low (high) temperature,
where the finite-$\mu$ corrections are pronounced (negligible).

\begin{figure*}[!htbp]
\begin{center}
\setlength{\abovecaptionskip}{-0.1mm}
\includegraphics[width=.47\textwidth]{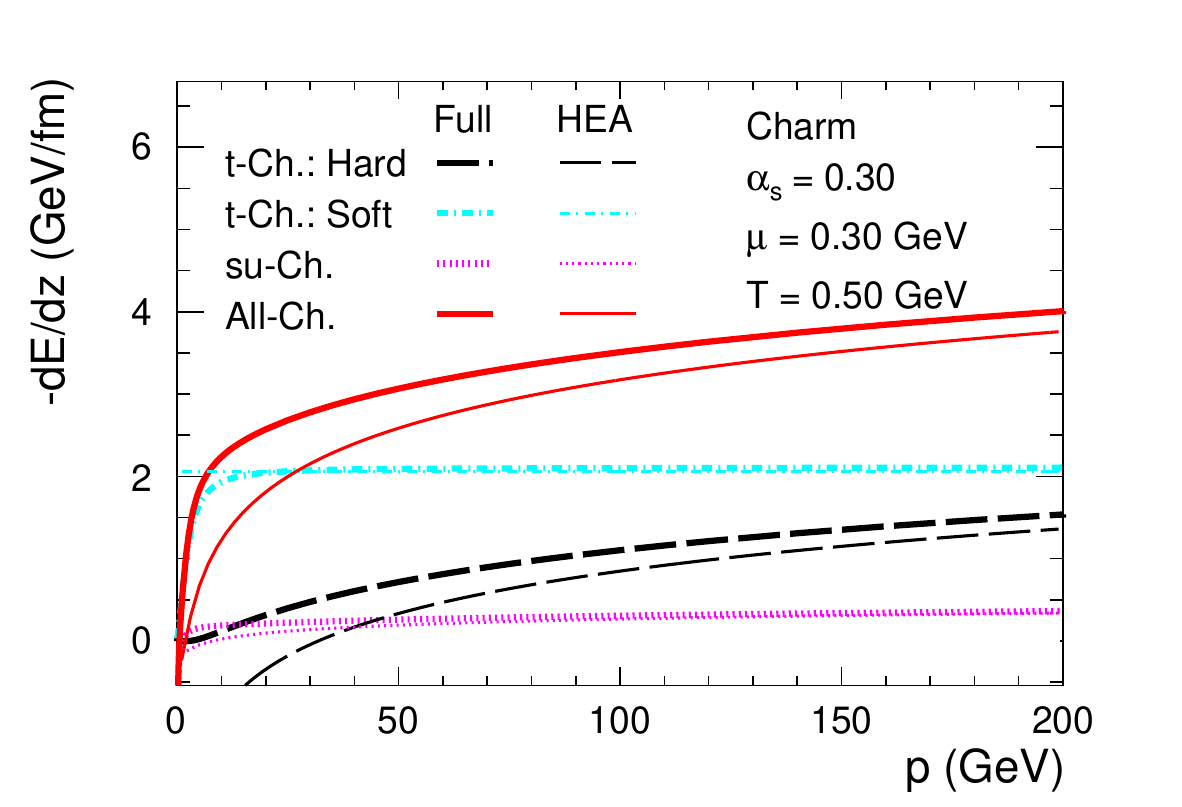}
\caption{(Color online) Comparison of charm quark energy loss based on
the full calculations [``Full''; thick curves; Eqs.~(\ref{eq:ELoss_Soft_vsX}), (\ref{eq:ELoss_Hard}) and (\ref{eq:dEdz_Hard_su_Perturbative})]
and the high-energy approximation
[``HEA''; thin curves; Eqs.~(\ref{eq:ELoss_Soft_HEA}), (\ref{eq:ELoss_Hard_Qq_HEA}), (\ref{eq:ELoss_Hard_Qg_t_HEA}) and (\ref{eq:ELoss_Hard_suCh_Def})]
as a function of heavy-quark momentum.
The associated results are shown as thick and thin curves, respectively.
Various contributions to the energy loss are displayed separately as curves with different styles.}
\label{fig:Charm_dEdz_FullHEA_variousChannels}
\end{center}
\end{figure*}
In Fig.~\ref{fig:Charm_dEdz_FullHEA_variousChannels}, charm quark $dE/dz$ evaluated
at fixed $\alpha_{s}=0.3$, $T=0.5~{\rm GeV}$ and $\mu=0.3~{\rm GeV}$,
are shown as a function of heavy-quark momentum.
The contributions from various channels are shown separately as curves with different styles.
By comparing the results based on the full calculations
[``Full''; thick curves; Eqs.~(\ref{eq:ELoss_Soft_vsX}), (\ref{eq:ELoss_Hard}) and (\ref{eq:dEdz_Hard_su_Perturbative})]
with that based on the high-energy approximation
[``HEA''; thin curves; Eqs.~(\ref{eq:ELoss_Soft_HEA}), (\ref{eq:ELoss_Hard_Qq_HEA}), (\ref{eq:ELoss_Hard_Qg_t_HEA}) and (\ref{eq:ELoss_Hard_suCh_Def})],
we find that (i) for each channel, the energy loss exhibits asymptotic growth at high momenta $p\gg m_{\rm Q}$,
while a sizable discrepancy is observed in the low and moderate momentum region;
(ii) for the soft component (dot-dashed cyan curves),
the $p$-dependence vanishes in the ``HEA'', as expected [Eq.~(\ref{eq:ELoss_Soft_HEA})].
The energy loss with the ``Full'' approach in each channel allows to
quantify the corresponding analytical result based on the ``HEA'',
in particular at high momentum region.
Moreover, it leaves room to investigate the charm quark energy loss
at low and moderate momentum $p\lesssim50~{\rm GeV}$.
Similar conclusions can be drawn for bottom quarks.

\begin{figure*}[!htbp]
\begin{center}
\setlength{\abovecaptionskip}{-0.1mm}
\includegraphics[width=.47\textwidth]{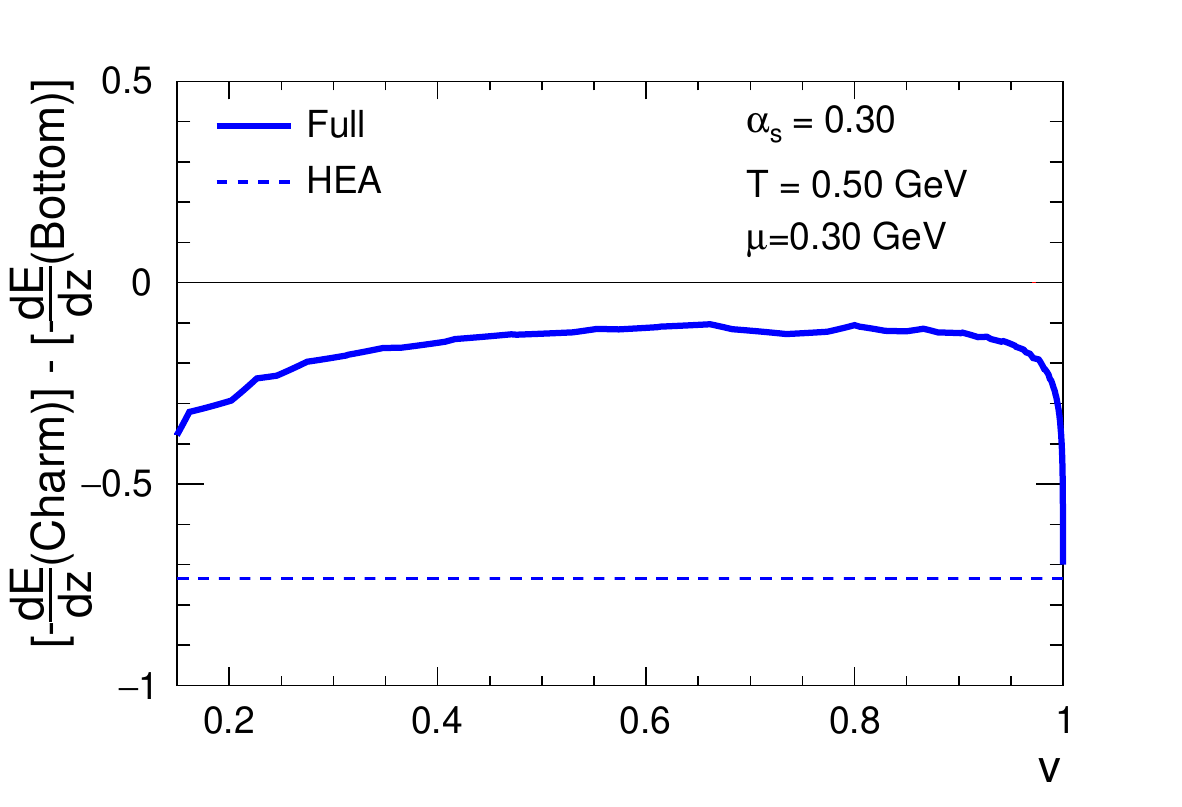}
\caption{(Color online) Comparison of $dE/dz$ for charm and bottom quarks
at fixed coupling $\alpha_{s}=0.3$, given temperature $T=0.5~{\rm GeV}$
and chemical potential $\mu=0.3~{\rm GeV}$.}
\label{fig:CharmBottomRatio_dEdz_vsV}
\end{center}
\end{figure*}
Figure~\ref{fig:CharmBottomRatio_dEdz_vsV} shows the discrepancy (=``Charm--Bottom'')
of the energy loss between charm and bottom quarks as a function of heavy-quark velocity.
The results obtained with the full calculation (``Full'') and the high-energy approximation (``HEA'')
are presented as solid and dashed curves, respectively.
This discrepancy in the ``Full'' calculation is negative over the entire velocity range,
indicating larger energy loss for the heavier quark.
The ``HEA'' result is independent of the heavy-quark velocity,
as for a given velocity it depends on terms involving the ratio $m_{c}/m_{b}$
rather than the velocity [see Eq.~(\ref{eq:ELoss_All_HEA})].
The discrepancy derived from the above two scenarios, as expected, converge for $v\rightarrow1$,
which corresponds to the ultrarelativistic limit.
These conclusions remain valid for the case of $\mu=0$~\cite{peng2024unraveling}.

\begin{figure*}[!htbp]
\begin{center}
\setlength{\abovecaptionskip}{-0.1mm}
\includegraphics[width=.48\textwidth]{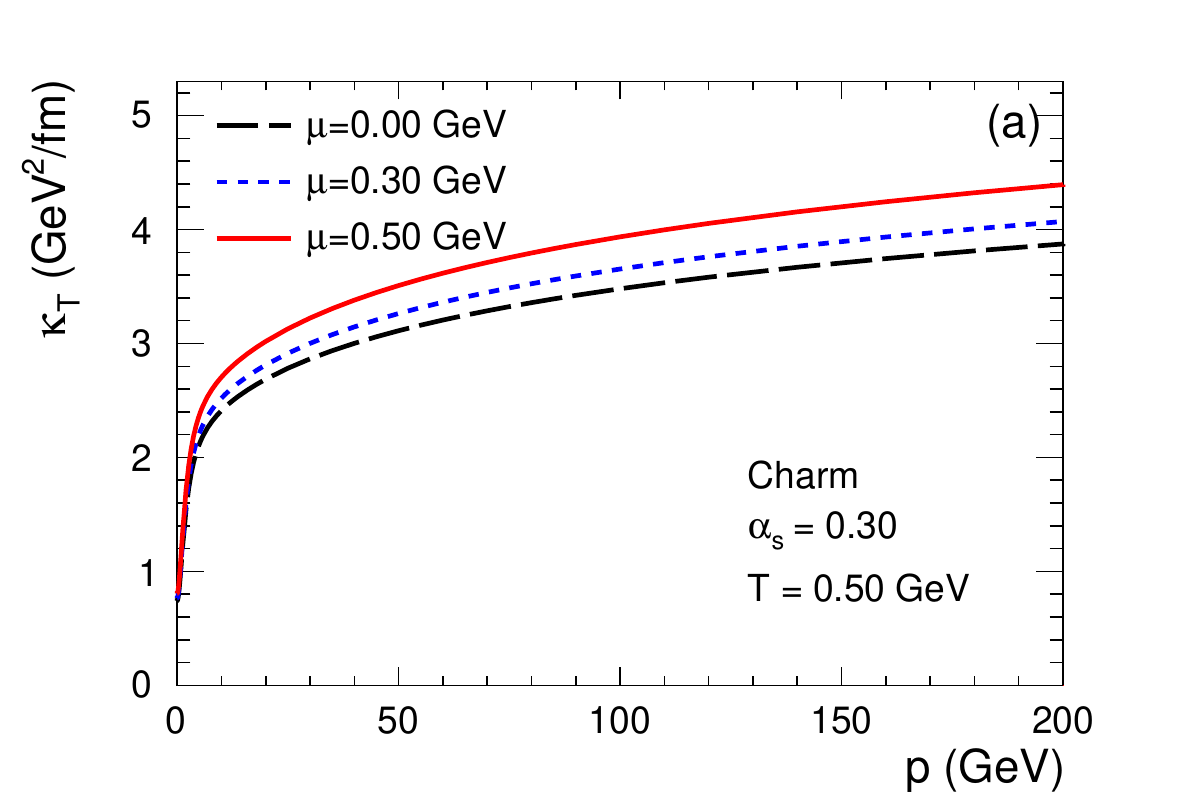}
\includegraphics[width=.48\textwidth]{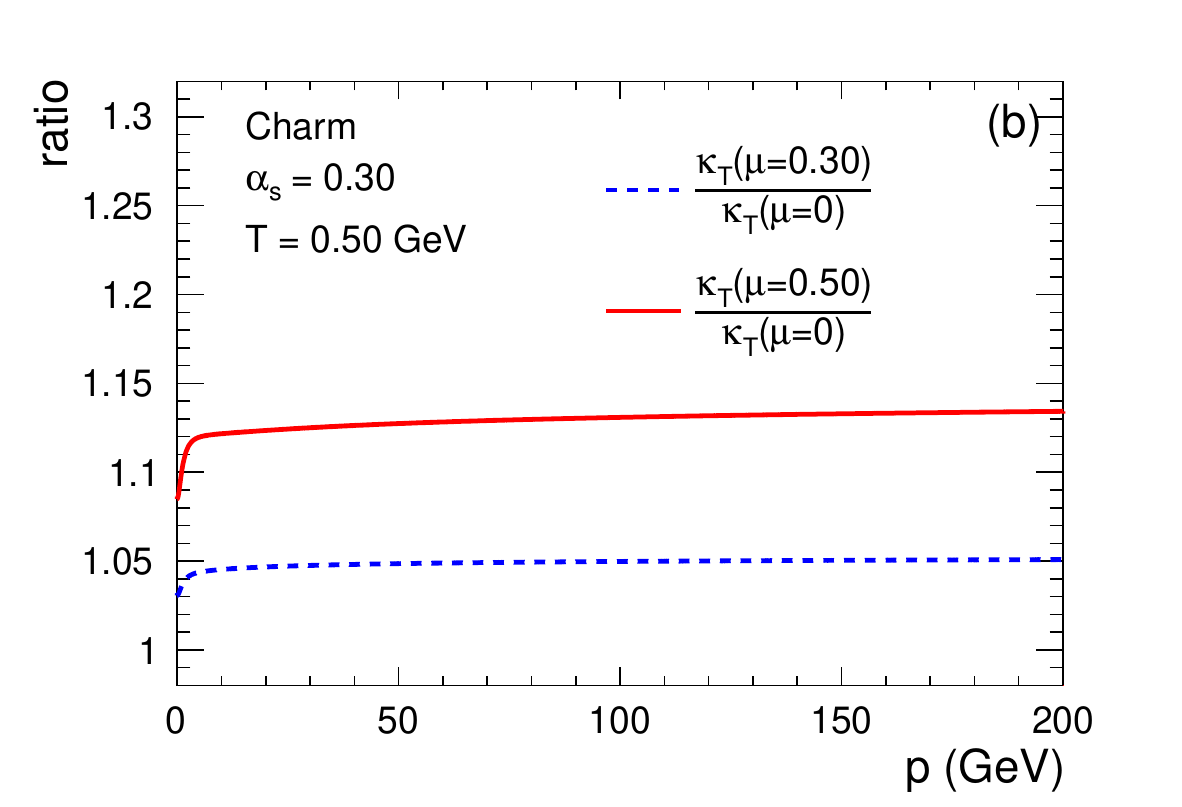}
\caption{(Color online) comparison of the transverse momentum diffusion coefficient $\kappa_{T}$ as a function of heavy-quark momentum,
for charm quark with $\alpha_{s}=0.3$, $-t^{\ast}=8M_{D}^{2}$,
$\mu=0$ (dotted curves), $\mu=0.3~{\rm GeV}$ (solid curves) and at a given temperature $T=0.5~{\rm GeV}$,
contributed by soft interactions in $t$-channel (pink curves),
and hard interactions in $t$-channel (blue curves),
$su$-channels (black curves).
The combined results, i.e. the contributions from the all components (red curves), are shown for comparison.
Right (b): comparison of $\kappa_{T}$ obtained for $\mu=0.3~{\rm GeV}$
with respect to that for $\mu=0$.}
\label{fig:Charm_KappaT_vsP_variousMu}
\end{center}
\end{figure*}
Figure~\ref{fig:Charm_KappaT_vsP_variousMu}(a) presents the $p$-dependence of the transverse momentum diffusion coefficient $\kappa_{T}$
of charm quarks for $\mu=0$ (dotted curves) and $\mu=0.3~{\rm GeV}$ (solid curves) at a given temperature $T=0.5~{\rm GeV}$.
As observed for $dE/dz$ (panel-a in Fig.~\ref{fig:Charm_dEdz_AllChan_variousMu_vsP}),
a monotonously rising $p$-dependence is observed for $\kappa_{T}$ with finite-$\mu$,
which is systematically larger than that with $\mu=0$.
Figure~\ref{fig:Charm_KappaT_vsP_variousMu}(b) shows the
ratios of $\kappa_{T}$ obtained at finite-$\mu$ with respect to that with $\mu=0$.
The ratios exhibit a visible $p$-dependence only in the range $p\lesssim3~{\rm GeV}$,
followed by an almost flat behavior toward higher momentum.
The enhancement of $\kappa_{T}$ with $\mu=0.5~\rm{GeV}$
reaches $\sim14\%$ at maximum in the considered momentum region.

\begin{figure*}[!htbp]
\begin{center}
\setlength{\abovecaptionskip}{-0.1mm}
\includegraphics[width=.48\textwidth]{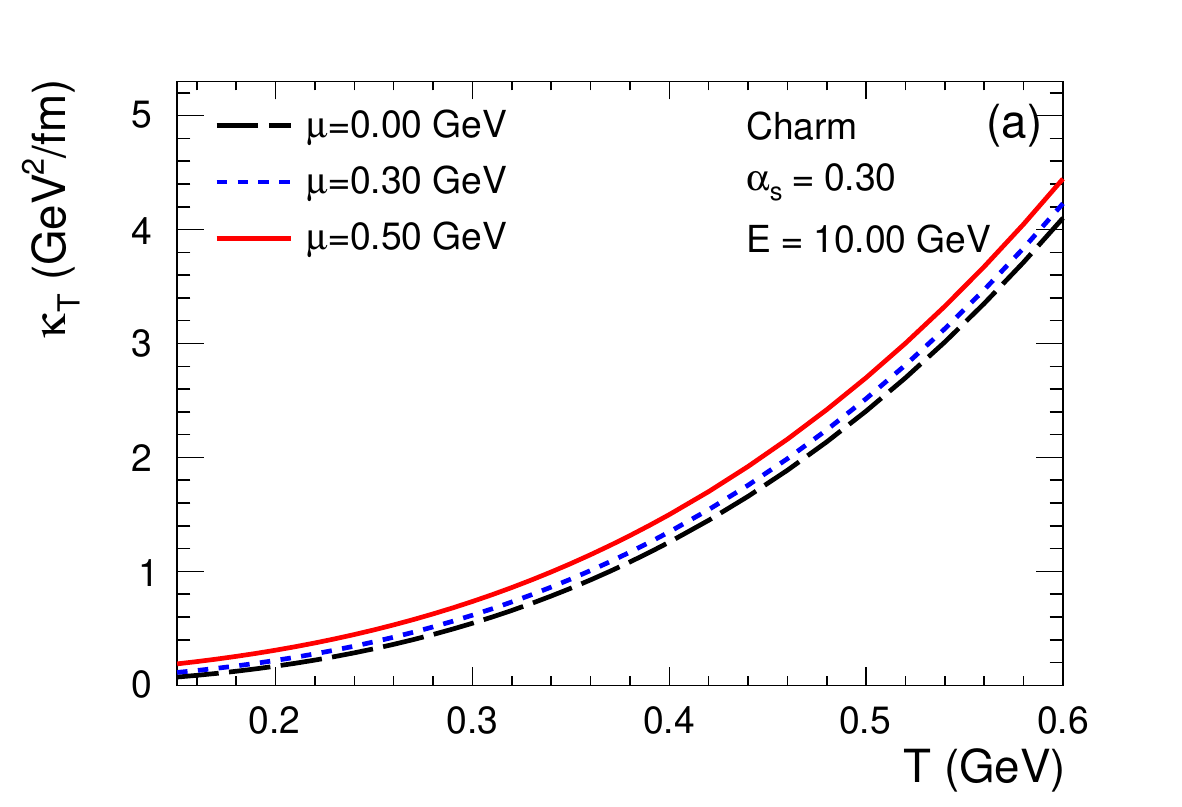}
\includegraphics[width=.48\textwidth]{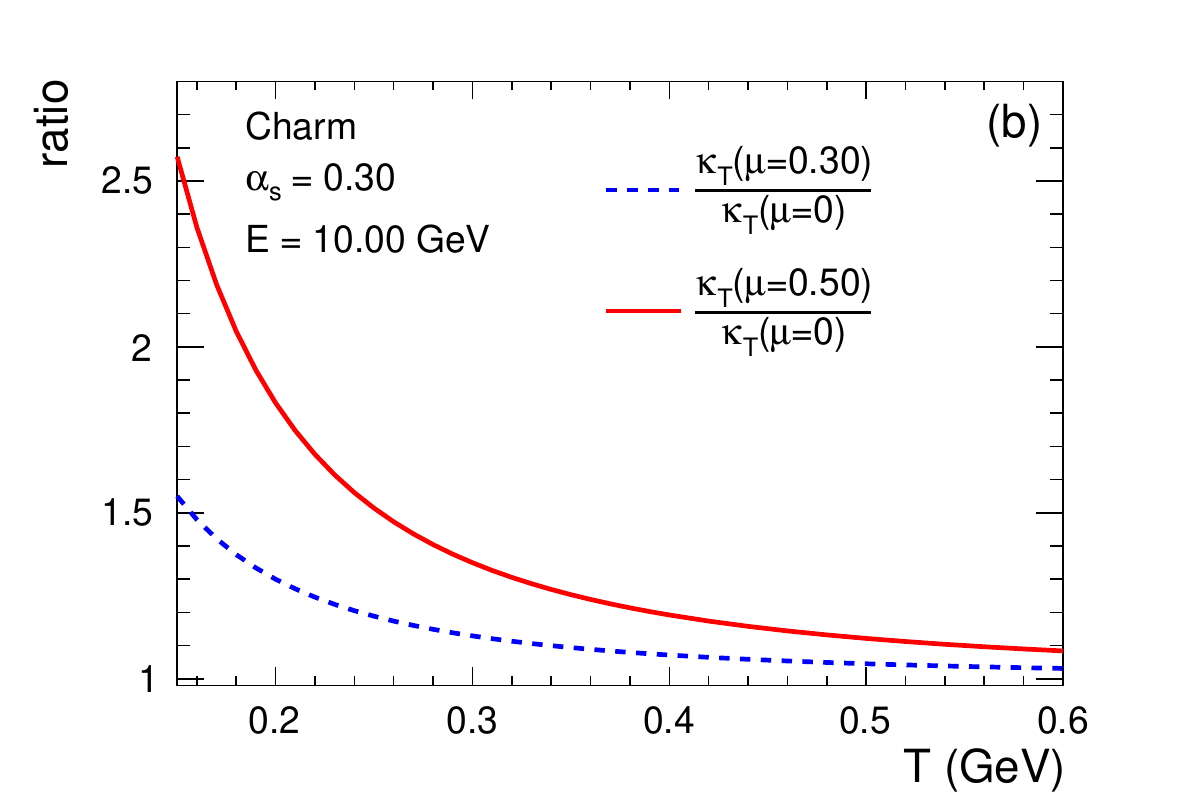}
\caption{Same as Fig.~\ref{fig:Charm_KappaT_vsP_variousMu} but as a function of temperature.}
\label{fig:Charm_KappaT_vsT_variousMu}
\end{center}
\end{figure*}
Figure~\ref{fig:Charm_KappaT_vsT_variousMu} presents the same results
as in Fig.~\ref{fig:Charm_KappaT_vsP_variousMu} but as a function of temperature.
The ratios, as observed in Fig.~\ref{fig:Charm_dEdz_AllChan_variousMu_vsT},
exhibit a strong (weak) $T$-dependence at low (high) temperature,
where the finite-$\mu$ corrections are pronounced (negligible).
Similar behaviors can be found for the longitudinal momentum coefficient $\kappa_{L}$.

\section{Summary}\label{sec:conclusions}
In this work, we have extended the recently developed soft-hard factorization model
for heavy-quark transport to finite chemical potential.
The $\mu$-dependence enters through two distinct channels:
the Debye screening mass $M_D(T,\mu)$ in the soft sector,
and the modified quark and antiquark distribution functions in the hard sector.
With this extended framework we have computed the collisional energy loss $-dE/dz$ and
the momentum-diffusion coefficients $\kappa_{T,L}$ of charm and bottom quarks
over a wide range of heavy-quark energies, medium temperatures and
chemical potentials relevant for baryon-rich QCD matter.

Our results show that both $-dE/dz$ and $\kappa_{T,L}$ increase significantly with $\mu$,
with the enhancement being most pronounced at low temperatures.
In the high-energy approximation the finite-$\mu$ corrections manifest as logarithmic terms:
a soft logarithm $\sim \mu^2 \ln(|t^{*}|/M_D^2)$ from scattering off thermal gluonic excitations,
and a hard logarithm $\sim \mu^2 \ln(E_1T/|t^{*}|)$ from scattering off thermal quarks.
The sum of these contributions is free of the separation scale $t^{*}$,
in agreement with general consistency requirements.
We also confirm the expected mass hierarchy $-dE/dz(charm)<-dE/dz(bottom)$ at fixed velocity.

These findings underscore the importance of including finite chemical potential
in theoretical descriptions of heavy-quark transport in baryon-rich environments
such as those created in the RHIC Beam Energy Scan,
and in the upcoming FAIR and NICA experiments.
The enhanced energy loss and diffusion coefficients predicted here
are essential for quantitative interpretation of heavy-flavor observables in these regimes.

We end with discussions on a few important caveats in the present study that deserve emphasis and that call for future investigations: 
\begin{itemize}
\item
The Debye screening mass $M_D(T,\mu)$ corresponds to the leading-order perturbative QCD result, which is quantitatively reliable at sufficiently high temperatures where the coupling remains weak. Near the critical temperature $T_c$, however, nonperturbative effects become increasingly important, and lattice QCD studies have revealed a more intricate dependence of $M_D$ on both $T$ and $\mu$ than the simple perturbative scaling suggests \cite{Bellwied:2015rza, Borsanyi:2020fev}. In this work, the perturbative form of $M_D(T,\mu)$ is therefore used as a controlled baseline that captures the qualitative trends of chemical-potential effects; future extensions incorporating lattice-extracted or effective-model-motivated screening masses will be essential for quantitative precision at the moderate-temperature regimes relevant for the RHIC Beam Energy Scan, FAIR, and NICA programs.
\item
In the present implementation, the medium partons participating in hard scatterings ($-t>-t^{\ast}$) are treated as massless and follow bare dispersion relations $E=|\vec{p}|$. Consequently, the chemical potential dependence enters the hard sector only through the fermionic thermal distributions $\mathcal{N}_{F}(E,T,\mu)$ in Eq.~(\ref{eq:ThermalDis_FermionAvg}), without introducing additional $\mu$-dependent thermal masses. This choice follows the theoretical logic of the soft-hard factorization framework~\cite{braaten1991calculation, PhysRevD.44.R2625}, in which all medium-induced screening and collective effects are resummed exclusively in the soft domain ($-t<-t^{\ast}$) through the HTL propagator and its $\mu$-dependent Debye mass $M_D(T,\mu)$. In contrast, dynamical quasiparticle models such as the DQPM~\cite{Berrehrah:2014tva} incorporate $\mu$-dependent parton masses, thereby introducing an additional entry point for baryon-density effects. Neglecting this contribution in the present work yields a conservative estimate of the $\mu$-driven enhancement of the transport coefficients; integrating such self-consistent quasiparticle effects into the SHFM will be an important direction for future refinement.
\item
Future work will also extend this framework to incorporate gluon radiation and absorption,
enabling a unified treatment of both elastic and inelastic interactions.
Such an extension will be crucial for a comprehensive description of heavy-flavor
dynamics and the extraction of QGP transport properties under extreme conditions.
\end{itemize}


\begin{acknowledgements}
This work is supported by the National Natural Science Foundation of China (NSFC)
under Grants No.12375137 and No.12005114.
\end{acknowledgements}



\appendix

\section{Derivation of the energy loss in soft region [Eq.~(\ref{eq:ELoss_Soft_vsX})]}\label{appendix:Derive_ELoss_Soft_vsX}
\setcounter{equation}{0}
\renewcommand\theequation{A\arabic{equation}}

The bare quark propagator $S(K)$ in Eq.~(\ref{eq:Sigma_Soft1}) is
\begin{equation}\label{eq:QuarkPropagator_Def}
D(K=P_1-Q) = \frac{1}{\slashed{K}-m_1} = \frac{\slashed{K}+m_1}{K^2-m_1^2}.
\end{equation}
Substituting Eqs.~(\ref{eq:GluonPrp_HTL}) and (\ref{eq:QuarkPropagator_Def}) into Eq.~(\ref{eq:Sigma_Soft1}),
the trace in Eq.~(\ref{eq:Weldon}) becomes:
\begin{equation}
\begin{aligned}\label{eq:Trace_1}
Tr[(\slashed{P}_{1}+m_1)\cdot\Sigma(P_1)] =& 4C_{F}g^{2}T\sum_{q^{0}}\int\frac{d^{3}\vec{q}}{(2\pi)^{3}}
\frac{1}{k_{0}^{2}-E_{k}^{2}}  \biggr\{ G_L\bigr[ (p_{1}^{0})^{2}+p_{1}^{2}-p_{1}^{0}q^{0}-\vec{p}_1\cdot\vec{q}+m_{1}^{2} \bigr] \\
& + 2G_T \bigr[ (p_{1}^{0})^{2}-p_{1}^{0}q^{0}+\vec{p}_1\cdot\vec{q}-(\vec{p}_1\cdot\widehat{q})^{2}-m_{1}^{2}\bigr] \biggr\},
\end{aligned}
\end{equation}
where $E_{k}=\sqrt{\vec{k}^{\;2}+m_{1}^{2}}$ is the energy of internal quark, and $\widehat{q}=\vec{q}/|\vec{q}\;|$.

The mixed representation of the HTL-resummed gluon propagator reads~\cite{li2021langevin}
\begin{equation}\label{eq:GluonPropagator_Mixed}
G_{T/L}(Q)=\int_{0}^{\frac{1}{T}}d\tau e^{q^{0}\tau}\int_{-\infty}^{+\infty}d\omega\frac{1}{2\pi}\rho_{T/L}[1+n_{B}(\omega,T)]e^{-\omega\tau},
\end{equation}
where $n_B(\omega,T)$ is the Bose-Einstein distribution [Eq.~(\ref{eq:ThermalDis_Boson})].
The spectral functions are defined by
\begin{equation}
\begin{aligned} \label{eq:BFET_GammaRhoTL1}
\rho_{T/L}(\omega,q,\mu) = 2\cdot Im G_{T/L} (\omega+i\eta,q),
\end{aligned}
\end{equation}
obtained by analytic continuation $q^{0}=\omega+i\eta \; (\eta\rightarrow 0^{+})$ in
Eqs.~(\ref{eq:Propagator_Soft_T}) and (\ref{eq:Propagator_Soft_L}):
\begin{align}
\rho_T(\omega,q,\mu) &= \frac{\pi \omega M_D^2}{2q^3}(q^2 - \omega^2) 
\left\{ \left[ q^2 - \omega^2 + \frac{\omega^2 M_D^2}{2q^2} \right] 
\times \left( 1 + \frac{q^2 - \omega^2}{2\omega q} \ln \frac{q + \omega}{q - \omega} \right)^2 
+ \left( \frac{\pi \omega M_D^2}{4q^3}(q^2 - \omega^2) \right)^2 \right\}^{-1}, \label{eq:SpecFunc_T_vsOmega} \\
\rho_L(\omega,q,\mu) &= \frac{\pi \omega M_D^2}{q} 
\left\{ \left[ q^2 + M_D^2 \left( 1 - \frac{\omega}{2q} \ln \frac{q + \omega}{q - \omega} \right) \right]^2 
+ \left( \frac{\pi \omega M_D^2}{2q} \right)^2 \right\}^{-1} \label{eq:SpecFunc_L_vsOmega}.
\end{align}
For the bare quark propagator, its spectral function is a delta function, and the mixed representation is
\begin{align}\label{eq:QuarkPropagator_Mixed}
\frac{1}{k_{0}^{2} - E_{k}^2} &= -\int_0^{1/T} d\tau^{\prime} \, e^{k_0 \tau^{\prime}} \frac{1}{2E_{k}}
\left\{ \left[1 - n_F^{+}(E_2,T,\mu)\right] e^{-E_{k} \tau^{\prime}} - n_F^{-}(E_2,T,\mu) e^{E_{k} \tau^{\prime}} \right\},
\end{align}
where $n_F^\pm(E_2,T,\mu)$ are the Fermi-Dirac distributions [Eq.~(\ref{eq:ThermalDis_FermionInd})].

Substituting Eqs.~(\ref{eq:GluonPropagator_Mixed}) and (\ref{eq:QuarkPropagator_Mixed}) into Eq.~(\ref{eq:Trace_1}),
and using the identities
\begin{subequations}
\begin{align}
\sum_{q^{0}} e^{q^{0} (\tau - \tau^{\prime})} &= \frac{1}{T} \delta(\tau - \tau^{\prime}), \\
\sum_{q^{0}} q^{0} e^{q^{0} (\tau - \tau^{\prime})} &= \frac{1}{T} \delta^{\prime}(\tau - \tau^{\prime}),
\end{align}
\end{subequations}
we perform the sums over $q_0$ and calculate, in turn, the integrals over $\tau$ and $\tau'$.
Following Ref.~\cite{braaten1991energy}, we set $e^{(p_1^0 - \mu)/T} = -1$
for $p_1^0 = i(2n+1)\pi T + \mu$ to eliminate exponentials,
then analytically continue to real Minkowski energy $p_1^0 = E_1 + i\epsilon$
as required in Eq.~(\ref{eq:Weldon}).
Taking the imaginary part of Eq.~(\ref{eq:Trace_1}) yields
\begin{equation}
\begin{aligned}\label{eq:Trace_2}
Tr[(\slashed{P}_{1}+m_1)\cdot Im\Sigma(E_1+i\epsilon,\vec{p}_1)] =& -C_{F}g^{2} [1-n_{F}^{+}(E_1,\mu)]^{-1} \frac{1}{E_{k}}
\int\frac{d^{3}\vec{q}}{(2\pi)^{3}} \int_{-\infty}^{\infty}d\omega [1+n_{B}(\omega,T)] \\
& \biggr\{ \bigr[1-n_{F}^+(E_2,T,\mu)\bigr]\delta(E_1-E_{k}-\omega) - n_F^{-}(E_2,T,\mu)\delta(E_1+E_{k}-\omega) \biggr\} \\
& \biggr\{ \rho_{L}\bigr[(p_{1}^{0})^{2}+p_{1}^{2}-p_{1}^{0}\omega-\vec{p}_{1}\cdot\vec{q}+m_{1}^{2}\bigr] + 2\rho_{T}\bigr[(p_{1}^{0})^{2}-p_{1}^{0}\omega+\vec{p}_{1}\cdot\vec{q}-(\vec{p}_{1}\cdot\widehat{q})^{2}-m_{1}^{2}) \bigr] \biggr\},
\end{aligned}
\end{equation}

To simplify, we adopt the following approximations~\cite{li2021langevin}:
\begin{enumerate}
\item[(1)] $n_{F}^{\pm}(E_1,T,\mu)\approx 0$ since the injected heavy quark energy is substantially larger than the medium temperature $E_{1}\gg T$;
\item[(2)] $\delta(E_1+E_{k}-\omega)=0$ since we focus on the low $\omega$ region $\omega\sim T\ll E_{1}\sim E_{k}$;
\item[(3)] $E_{k}\approx E_{1}-\vec{v} \cdot \vec{q}$ is valid for collinear scattering processes with small momentum transfer.
\end{enumerate}
Equation~(\ref{eq:Trace_2}) then reduces to
\begin{equation}
\begin{aligned}\label{eq:Trace_3}
Tr[(\slashed{P}_{1}+m_1)\cdot Im\Sigma(E_1+i\epsilon,\vec{p}_1)]
=& -2C_{F}g^{2}E_{1} [1-n_{F}^{+}(E_1,T,\mu)]^{-1} \int \frac{d^3 \vec{q}}{(2\pi)^3} \\
& \int_{-\infty}^{\infty} d\omega [1+n_{B}(\omega,T)] \delta(\omega - \vec{v}_{1}\cdot\vec{q}\;) 
\left\{ \rho_L + \vec{v}_{1}^{\;2} \left[1 - (\widehat{v}_{1}\cdot\widehat{q})^2 \right]\rho_T \right\}.
\end{aligned}
\end{equation}
Substituting Eq.~(\ref{eq:Trace_3}) back into Eq.~(\ref{eq:Weldon}),
the soft collision rate becomes
\begin{align}
\Gamma_{(t)}^{soft}(E_1,T,\mu) &= C_F g^2 \int \frac{d^3\vec{q}}{(2\pi)^3} \int d\omega
[1+n_{B}(\omega,T)] \delta(\omega - \vec{v}_1 \cdot \vec{q})
\left\{ \rho_L + \vec{v}^2 \left[1 - (\widehat{v}\cdot\widehat{q})^2\right] \rho_T \right\},
\end{align}
which matches the result in Refs.~\cite{li2021langevin,peng2024unraveling}
but with the Debye mass at finite $\mu$.
The energy loss is
\begin{equation}
\begin{aligned}\label{eq:ELoss_Soft_vsOmega}
\left[-{\frac{dE}{dz}}\right]_{(t)}^{soft}(E_1,T,\mu) &= \frac{C_{F}g^{2}}{4\pi^{2}v_{1}^{2}}\int_{0}^{q_{max}}dq \; q
\int_{-vq}^{vq}d\omega \; \omega [1+n_{B}(\omega,T)]
\left[\rho_{L}(\omega,q,\mu)+v_{1}^{2}\left(1-\frac{\omega^{2}}{v_{1}^{2}q^{2}}\right)\rho_{T}(\omega,q,\mu)\right],
\end{aligned}
\end{equation}
where $\vec{v}_1 = \vec{p}_1/E_1$ is the heavy quark velocity,
and $q_{\max} = \sqrt{4E_{1}T}$ in the high-energy limit~\cite{bjorken1982energy}.

Introducing the variable transformation:
\begin{equation}
\begin{aligned}
	&t = \omega^2 - q^2 < 0, \quad && q^2 = \frac{-t}{1-x^2}, \\
	&x = \frac{\omega}{q} < v, \quad && \omega = x\sqrt{\frac{-t}{1-x^2}},
\end{aligned}
\label{eq:variable_change}
\end{equation}
with Jacobian
\begin{equation}
dtdx = \left| \frac{\partial (t,x)}{\partial (q,\omega)} \right| dq  d\omega = 2(1-x^2)  dq  d\omega,
\end{equation}
we can rewrite Eq.~(\ref{eq:ELoss_Soft_vsOmega}) as Eq.~(\ref{eq:ELoss_Soft_vsX}), with
\begin{align}
\rho_T(t,x,\mu) &= \frac{\pi M_D^2}{2}x(1-x^2)\left\{\left[-t+\frac{M_D^2}{2}x^2\left(1+\frac{1-x^2}{2x}\ln\frac{1+x}{1-x}\right)\right]^2+\left(\frac{\pi M_D^2}{4}x(1-x^2)\right)^2\right\}^{-1}, \label{eq:SpecFunc_T_vsX} \\
\rho_L(t,x,\mu) &= \pi M_D^2x\left\{\left[\frac{-t}{1-x^2}+M_D^2\left(1-\frac{x}{2}\ln\frac{1+x}{1-x}\right)\right]^2+\left(\frac{\pi M_D^2}{2}x\right)^2\right\}^{-1}. \label{eq:SpecFunc_L_vsX}
\end{align}

\input{ChemPotential.bbl}
%
\end{document}

%% file: ChemPotential.bbl
%

%% file: ChemPotential.bbl
\begin{thebibliography}{53}%
\makeatletter
\providecommand \@ifxundefined [1]{%
 \@ifx{#1\undefined}
}%
\providecommand \@ifnum [1]{%
 \ifnum #1\expandafter \@firstoftwo
 \else \expandafter \@secondoftwo
 \fi
}%
\providecommand \@ifx [1]{%
 \ifx #1\expandafter \@firstoftwo
 \else \expandafter \@secondoftwo
 \fi
}%
\providecommand \natexlab [1]{#1}%
\providecommand \enquote  [1]{``#1''}%
\providecommand \bibnamefont  [1]{#1}%
\providecommand \bibfnamefont [1]{#1}%
\providecommand \citenamefont [1]{#1}%
\providecommand \href@noop [0]{\@secondoftwo}%
\providecommand \href [0]{\begingroup \@sanitize@url \@href}%
\providecommand \@href[1]{\@@startlink{#1}\@@href}%
\providecommand \@@href[1]{\endgroup#1\@@endlink}%
\providecommand \@sanitize@url [0]{\catcode `\\12\catcode `\$12\catcode
  `\&12\catcode `\#12\catcode `\^12\catcode `\_12\catcode `\%12\relax}%
\providecommand \@@startlink[1]{}%
\providecommand \@@endlink[0]{}%
\providecommand \url  [0]{\begingroup\@sanitize@url \@url }%
\providecommand \@url [1]{\endgroup\@href {#1}{\urlprefix }}%
\providecommand \urlprefix  [0]{URL }%
\providecommand \Eprint [0]{\href }%
\providecommand \doibase [0]{http://dx.doi.org/}%
\providecommand \selectlanguage [0]{\@gobble}%
\providecommand \bibinfo  [0]{\@secondoftwo}%
\providecommand \bibfield  [0]{\@secondoftwo}%
\providecommand \translation [1]{[#1]}%
\providecommand \BibitemOpen [0]{}%
\providecommand \bibitemStop [0]{}%
\providecommand \bibitemNoStop [0]{.\EOS\space}%
\providecommand \EOS [0]{\spacefactor3000\relax}%
\providecommand \BibitemShut  [1]{\csname bibitem#1\endcsname}%
\let\auto@bib@innerbib\@empty
\bibitem [{\citenamefont {Gyulassy}\ and\ \citenamefont
  {McLerran}(2005)}]{Gyulassy:2004zy}%
  \BibitemOpen
  \bibfield  {author} {\bibinfo {author} {\bibfnamefont {M.}~\bibnamefont
  {Gyulassy}}\ and\ \bibinfo {author} {\bibfnamefont {L.}~\bibnamefont
  {McLerran}},\ }\href {\doibase 10.1016/j.nuclphysa.2004.10.034} {\bibfield
  {journal} {\bibinfo  {journal} {Nucl. Phys. A}\ }\textbf {\bibinfo {volume}
  {750}},\ \bibinfo {pages} {30} (\bibinfo {year} {2005})},\ \Eprint
  {http://arxiv.org/abs/nucl-th/0405013} {arXiv:nucl-th/0405013} \BibitemShut
  {NoStop}%
\bibitem [{\citenamefont {Shuryak}(2005)}]{Shuryak:2004cy}%
  \BibitemOpen
  \bibfield  {author} {\bibinfo {author} {\bibfnamefont {E.~V.}\ \bibnamefont
  {Shuryak}},\ }\href {\doibase 10.1016/j.nuclphysa.2004.10.022} {\bibfield
  {journal} {\bibinfo  {journal} {Nucl. Phys. A}\ }\textbf {\bibinfo {volume}
  {750}},\ \bibinfo {pages} {64} (\bibinfo {year} {2005})},\ \Eprint
  {http://arxiv.org/abs/hep-ph/0405066} {arXiv:hep-ph/0405066} \BibitemShut
  {NoStop}%
\bibitem [{\citenamefont {{STAR Collaboration}}(2005)}]{STAR05}%
  \BibitemOpen
  \bibfield  {author} {\bibinfo {author} {\bibnamefont {{STAR
  Collaboration}}},\ }\href {\doibase 10.1016/j.nuclphysa.2005.03.085}
  {\bibfield  {journal} {\bibinfo  {journal} {Nucl.~Phys.~A}\ }\textbf
  {\bibinfo {volume} {757}},\ \bibinfo {pages} {102} (\bibinfo {year}
  {2005})}\BibitemShut {NoStop}%
\bibitem [{\citenamefont {Frawley}\ \emph {et~al.}(2008)\citenamefont
  {Frawley}, \citenamefont {Ullrich},\ and\ \citenamefont
  {Vogt}}]{Frawley:2008kk}%
  \BibitemOpen
  \bibfield  {author} {\bibinfo {author} {\bibfnamefont {A.~D.}\ \bibnamefont
  {Frawley}}, \bibinfo {author} {\bibfnamefont {T.}~\bibnamefont {Ullrich}}, \
  and\ \bibinfo {author} {\bibfnamefont {R.}~\bibnamefont {Vogt}},\ }\href
  {\doibase 10.1016/j.physrep.2008.04.002} {\bibfield  {journal} {\bibinfo
  {journal} {Phys. Rept.}\ }\textbf {\bibinfo {volume} {462}},\ \bibinfo
  {pages} {125} (\bibinfo {year} {2008})},\ \Eprint
  {http://arxiv.org/abs/0806.1013} {arXiv:0806.1013 [nucl-ex]} \BibitemShut
  {NoStop}%
\bibitem [{\citenamefont {Bzdak}\ \emph {et~al.}(2020)\citenamefont {Bzdak},
  \citenamefont {Esumi}, \citenamefont {Koch}, \citenamefont {Liao},
  \citenamefont {Stephanov},\ and\ \citenamefont {Xu}}]{Bzdak:2019pkr}%
  \BibitemOpen
  \bibfield  {author} {\bibinfo {author} {\bibfnamefont {A.}~\bibnamefont
  {Bzdak}}, \bibinfo {author} {\bibfnamefont {S.}~\bibnamefont {Esumi}},
  \bibinfo {author} {\bibfnamefont {V.}~\bibnamefont {Koch}}, \bibinfo {author}
  {\bibfnamefont {J.}~\bibnamefont {Liao}}, \bibinfo {author} {\bibfnamefont
  {M.}~\bibnamefont {Stephanov}}, \ and\ \bibinfo {author} {\bibfnamefont
  {N.}~\bibnamefont {Xu}},\ }\href {\doibase 10.1016/j.physrep.2020.01.005}
  {\bibfield  {journal} {\bibinfo  {journal} {Phys. Rept.}\ }\textbf {\bibinfo
  {volume} {853}},\ \bibinfo {pages} {1} (\bibinfo {year} {2020})},\ \Eprint
  {http://arxiv.org/abs/1906.00936} {arXiv:1906.00936 [nucl-th]} \BibitemShut
  {NoStop}%
\bibitem [{\citenamefont {Dong}\ \emph {et~al.}(2019)\citenamefont {Dong},
  \citenamefont {Lee},\ and\ \citenamefont {Rapp}}]{Dong:2019byy}%
  \BibitemOpen
  \bibfield  {author} {\bibinfo {author} {\bibfnamefont {X.}~\bibnamefont
  {Dong}}, \bibinfo {author} {\bibfnamefont {Y.}~\bibnamefont {Lee}}, \ and\
  \bibinfo {author} {\bibfnamefont {R.}~\bibnamefont {Rapp}},\ }\href {\doibase
  10.1146/annurev-nucl-101918-023806} {\bibfield  {journal} {\bibinfo
  {journal} {Ann. Rev. Nucl. Part. Sci.}\ }\textbf {\bibinfo {volume} {69}},\
  \bibinfo {pages} {417} (\bibinfo {year} {2019})},\ \Eprint
  {http://arxiv.org/abs/1903.07709} {arXiv:1903.07709 [nucl-ex]} \BibitemShut
  {NoStop}%
\bibitem [{\citenamefont {Tang}\ \emph {et~al.}(2020)\citenamefont {Tang},
  \citenamefont {Tang}, \citenamefont {Zha}, \citenamefont {Zha}, \citenamefont
  {Zhang},\ and\ \citenamefont {Zhang}}]{Tang:2020ame}%
  \BibitemOpen
  \bibfield  {author} {\bibinfo {author} {\bibfnamefont {Z.}~\bibnamefont
  {Tang}}, \bibinfo {author} {\bibfnamefont {Z.-B.}\ \bibnamefont {Tang}},
  \bibinfo {author} {\bibfnamefont {W.}~\bibnamefont {Zha}}, \bibinfo {author}
  {\bibfnamefont {W.-M.}\ \bibnamefont {Zha}}, \bibinfo {author} {\bibfnamefont
  {Y.}~\bibnamefont {Zhang}}, \ and\ \bibinfo {author} {\bibfnamefont {Y.-F.}\
  \bibnamefont {Zhang}},\ }\href {\doibase 10.1007/s41365-020-00785-8}
  {\bibfield  {journal} {\bibinfo  {journal} {Nucl. Sci. Tech.}\ }\textbf
  {\bibinfo {volume} {31}},\ \bibinfo {pages} {81} (\bibinfo {year} {2020})},\
  \Eprint {http://arxiv.org/abs/2105.11656} {arXiv:2105.11656 [nucl-ex]}
  \BibitemShut {NoStop}%
\bibitem [{\citenamefont {Chen}\ \emph {et~al.}(2024)\citenamefont {Chen} \emph
  {et~al.}}]{Chen:2024aom}%
  \BibitemOpen
  \bibfield  {author} {\bibinfo {author} {\bibfnamefont {J.}~\bibnamefont
  {Chen}} \emph {et~al.},\ }\href {\doibase 10.1007/s41365-024-01591-2}
  {\bibfield  {journal} {\bibinfo  {journal} {Nucl. Sci. Tech.}\ }\textbf
  {\bibinfo {volume} {35}},\ \bibinfo {pages} {214} (\bibinfo {year} {2024})},\
  \Eprint {http://arxiv.org/abs/2407.02935} {arXiv:2407.02935 [nucl-ex]}
  \BibitemShut {NoStop}%
\bibitem [{\citenamefont {Acharya}\ \emph {et~al.}(2022)\citenamefont {Acharya}
  \emph {et~al.}}]{ALICE:2021rxa}%
  \BibitemOpen
  \bibfield  {author} {\bibinfo {author} {\bibfnamefont {S.}~\bibnamefont
  {Acharya}} \emph {et~al.} (\bibinfo {collaboration} {ALICE Collaboration}),\
  }\href {\doibase 10.1007/JHEP01(2022)174} {\bibfield  {journal} {\bibinfo
  {journal} {JHEP}\ }\textbf {\bibinfo {volume} {01}},\ \bibinfo {pages} {174}
  (\bibinfo {year} {2022})},\ \Eprint {http://arxiv.org/abs/2110.09420}
  {arXiv:2110.09420 [nucl-ex]} \BibitemShut {NoStop}%
\bibitem [{\citenamefont {Acharya}\ \emph {et~al.}(2024)\citenamefont {Acharya}
  \emph {et~al.}}]{ALICE:2022wpn}%
  \BibitemOpen
  \bibfield  {author} {\bibinfo {author} {\bibfnamefont {S.}~\bibnamefont
  {Acharya}} \emph {et~al.} (\bibinfo {collaboration} {ALICE Collaboration}),\
  }\href {\doibase 10.1140/epjc/s10052-024-12935-y} {\bibfield  {journal}
  {\bibinfo  {journal} {Eur. Phys. J. C}\ }\textbf {\bibinfo {volume} {84}},\
  \bibinfo {pages} {813} (\bibinfo {year} {2024})},\ \Eprint
  {http://arxiv.org/abs/2211.04384} {arXiv:2211.04384 [nucl-ex]} \BibitemShut
  {NoStop}%
\bibitem [{\citenamefont {Ablyazimov}\ \emph {et~al.}(2017)\citenamefont
  {Ablyazimov} \emph {et~al.}}]{CBM:2016kpk}%
  \BibitemOpen
  \bibfield  {author} {\bibinfo {author} {\bibfnamefont {T.}~\bibnamefont
  {Ablyazimov}} \emph {et~al.} (\bibinfo {collaboration} {CBM Collaboration}),\
  }\href {\doibase 10.1140/epja/i2017-12248-y} {\bibfield  {journal} {\bibinfo
  {journal} {Eur. Phys. J. A}\ }\textbf {\bibinfo {volume} {53}},\ \bibinfo
  {pages} {60} (\bibinfo {year} {2017})},\ \Eprint
  {http://arxiv.org/abs/1607.01487} {arXiv:1607.01487 [nucl-ex]} \BibitemShut
  {NoStop}%
\bibitem [{\citenamefont {Abgaryan}\ \emph {et~al.}(2022)\citenamefont
  {Abgaryan} \emph {et~al.}}]{MPD:2022qhn}%
  \BibitemOpen
  \bibfield  {author} {\bibinfo {author} {\bibfnamefont {V.}~\bibnamefont
  {Abgaryan}} \emph {et~al.} (\bibinfo {collaboration} {MPD Collaboration}),\
  }\href {\doibase 10.1140/epja/s10050-022-00750-6} {\bibfield  {journal}
  {\bibinfo  {journal} {Eur. Phys. J. A}\ }\textbf {\bibinfo {volume} {58}},\
  \bibinfo {pages} {140} (\bibinfo {year} {2022})},\ \Eprint
  {http://arxiv.org/abs/2202.08970} {arXiv:2202.08970 [physics.ins-det]}
  \BibitemShut {NoStop}%
\bibitem [{\citenamefont {Stephanov}\ \emph {et~al.}(1998)\citenamefont
  {Stephanov}, \citenamefont {Rajagopal},\ and\ \citenamefont
  {Shuryak}}]{Stephanov:1998dy}%
  \BibitemOpen
  \bibfield  {author} {\bibinfo {author} {\bibfnamefont {M.~A.}\ \bibnamefont
  {Stephanov}}, \bibinfo {author} {\bibfnamefont {K.}~\bibnamefont
  {Rajagopal}}, \ and\ \bibinfo {author} {\bibfnamefont {E.~V.}\ \bibnamefont
  {Shuryak}},\ }\href {\doibase 10.1103/PhysRevLett.81.4816} {\bibfield
  {journal} {\bibinfo  {journal} {Phys. Rev. Lett.}\ }\textbf {\bibinfo
  {volume} {81}},\ \bibinfo {pages} {4816} (\bibinfo {year} {1998})},\ \Eprint
  {http://arxiv.org/abs/hep-ph/9806219} {arXiv:hep-ph/9806219} \BibitemShut
  {NoStop}%
\bibitem [{\citenamefont {Fodor}\ and\ \citenamefont
  {Katz}(2004)}]{Fodor:2004nz}%
  \BibitemOpen
  \bibfield  {author} {\bibinfo {author} {\bibfnamefont {Z.}~\bibnamefont
  {Fodor}}\ and\ \bibinfo {author} {\bibfnamefont {S.~D.}\ \bibnamefont
  {Katz}},\ }\href {\doibase 10.1088/1126-6708/2004/04/050} {\bibfield
  {journal} {\bibinfo  {journal} {JHEP}\ }\textbf {\bibinfo {volume} {04}},\
  \bibinfo {pages} {050} (\bibinfo {year} {2004})},\ \Eprint
  {http://arxiv.org/abs/hep-lat/0402006} {arXiv:hep-lat/0402006} \BibitemShut
  {NoStop}%
\bibitem [{\citenamefont {{F.~Prino and R.~Rapp}}(2016)}]{RalfSummary16}%
  \BibitemOpen
  \bibfield  {author} {\bibinfo {author} {\bibnamefont {{F.~Prino and
  R.~Rapp}}},\ }\href {\doibase 10.1088/0954-3899/43/9/093002} {\bibfield
  {journal} {\bibinfo  {journal} {J.~Phys.~G}\ }\textbf {\bibinfo {volume}
  {43}},\ \bibinfo {pages} {093002} (\bibinfo {year} {2016})}\BibitemShut
  {NoStop}%
\bibitem [{\citenamefont {Chen}\ \emph {et~al.}(2022)\citenamefont {Chen},
  \citenamefont {Jiang}, \citenamefont {Liu}, \citenamefont {Liu},\ and\
  \citenamefont {Zhao}}]{Chen:2021akx}%
  \BibitemOpen
  \bibfield  {author} {\bibinfo {author} {\bibfnamefont {B.}~\bibnamefont
  {Chen}}, \bibinfo {author} {\bibfnamefont {L.}~\bibnamefont {Jiang}},
  \bibinfo {author} {\bibfnamefont {X.-H.}\ \bibnamefont {Liu}}, \bibinfo
  {author} {\bibfnamefont {Y.}~\bibnamefont {Liu}}, \ and\ \bibinfo {author}
  {\bibfnamefont {J.}~\bibnamefont {Zhao}},\ }\href {\doibase
  10.1103/PhysRevC.105.054901} {\bibfield  {journal} {\bibinfo  {journal}
  {Phys. Rev. C}\ }\textbf {\bibinfo {volume} {105}},\ \bibinfo {pages}
  {054901} (\bibinfo {year} {2022})},\ \Eprint
  {http://arxiv.org/abs/2107.00969} {arXiv:2107.00969 [hep-ph]} \BibitemShut
  {NoStop}%
\bibitem [{\citenamefont {He}\ \emph {et~al.}(2023)\citenamefont {He},
  \citenamefont {van Hees},\ and\ \citenamefont {Rapp}}]{He:2022ywp}%
  \BibitemOpen
  \bibfield  {author} {\bibinfo {author} {\bibfnamefont {M.}~\bibnamefont
  {He}}, \bibinfo {author} {\bibfnamefont {H.}~\bibnamefont {van Hees}}, \ and\
  \bibinfo {author} {\bibfnamefont {R.}~\bibnamefont {Rapp}},\ }\href {\doibase
  10.1016/j.ppnp.2023.104020} {\bibfield  {journal} {\bibinfo  {journal} {Prog.
  Part. Nucl. Phys.}\ }\textbf {\bibinfo {volume} {130}},\ \bibinfo {pages}
  {104020} (\bibinfo {year} {2023})},\ \Eprint
  {http://arxiv.org/abs/2204.09299} {arXiv:2204.09299 [hep-ph]} \BibitemShut
  {NoStop}%
\bibitem [{\citenamefont {Moore}\ and\ \citenamefont
  {Teaney}(2005)}]{PhysRevC.71.064904}%
  \BibitemOpen
  \bibfield  {author} {\bibinfo {author} {\bibfnamefont {G.~D.}\ \bibnamefont
  {Moore}}\ and\ \bibinfo {author} {\bibfnamefont {D.}~\bibnamefont {Teaney}},\
  }\href {\doibase 10.1103/PhysRevC.71.064904} {\bibfield  {journal} {\bibinfo
  {journal} {Phys. Rev. C}\ }\textbf {\bibinfo {volume} {71}},\ \bibinfo
  {pages} {064904} (\bibinfo {year} {2005})}\BibitemShut {NoStop}%
\bibitem [{\citenamefont {Alberico}\ \emph {et~al.}(2011)\citenamefont
  {Alberico}, \citenamefont {Beraudo}, \citenamefont {De~Pace}, \citenamefont
  {Molinari}, \citenamefont {Monteno}, \citenamefont {Nardi},\ and\
  \citenamefont {Prino}}]{Alberico:2011zy}%
  \BibitemOpen
  \bibfield  {author} {\bibinfo {author} {\bibfnamefont {W.~M.}\ \bibnamefont
  {Alberico}}, \bibinfo {author} {\bibfnamefont {A.}~\bibnamefont {Beraudo}},
  \bibinfo {author} {\bibfnamefont {A.}~\bibnamefont {De~Pace}}, \bibinfo
  {author} {\bibfnamefont {A.}~\bibnamefont {Molinari}}, \bibinfo {author}
  {\bibfnamefont {M.}~\bibnamefont {Monteno}}, \bibinfo {author} {\bibfnamefont
  {M.}~\bibnamefont {Nardi}}, \ and\ \bibinfo {author} {\bibfnamefont
  {F.}~\bibnamefont {Prino}},\ }\href {\doibase 10.1140/epjc/s10052-011-1666-6}
  {\bibfield  {journal} {\bibinfo  {journal} {Eur. Phys. J. C}\ }\textbf
  {\bibinfo {volume} {71}},\ \bibinfo {pages} {1666} (\bibinfo {year}
  {2011})},\ \Eprint {http://arxiv.org/abs/1101.6008} {arXiv:1101.6008
  [hep-ph]} \BibitemShut {NoStop}%
\bibitem [{\citenamefont {He}\ \emph {et~al.}(2013)\citenamefont {He},
  \citenamefont {Fries},\ and\ \citenamefont {Rapp}}]{PhysRevLett.110.112301}%
  \BibitemOpen
  \bibfield  {author} {\bibinfo {author} {\bibfnamefont {M.}~\bibnamefont
  {He}}, \bibinfo {author} {\bibfnamefont {R.~J.}\ \bibnamefont {Fries}}, \
  and\ \bibinfo {author} {\bibfnamefont {R.}~\bibnamefont {Rapp}},\ }\href
  {\doibase 10.1103/PhysRevLett.110.112301} {\bibfield  {journal} {\bibinfo
  {journal} {Phys. Rev. Lett.}\ }\textbf {\bibinfo {volume} {110}},\ \bibinfo
  {pages} {112301} (\bibinfo {year} {2013})}\BibitemShut {NoStop}%
\bibitem [{\citenamefont {Song}\ \emph {et~al.}(2016)\citenamefont {Song},
  \citenamefont {Berrehrah}, \citenamefont {Cabrera}, \citenamefont {Cassing},\
  and\ \citenamefont {Bratkovskaya}}]{PhysRevC.93.034906}%
  \BibitemOpen
  \bibfield  {author} {\bibinfo {author} {\bibfnamefont {T.}~\bibnamefont
  {Song}}, \bibinfo {author} {\bibfnamefont {H.}~\bibnamefont {Berrehrah}},
  \bibinfo {author} {\bibfnamefont {D.}~\bibnamefont {Cabrera}}, \bibinfo
  {author} {\bibfnamefont {W.}~\bibnamefont {Cassing}}, \ and\ \bibinfo
  {author} {\bibfnamefont {E.}~\bibnamefont {Bratkovskaya}},\ }\href {\doibase
  10.1103/PhysRevC.93.034906} {\bibfield  {journal} {\bibinfo  {journal} {Phys.
  Rev. C}\ }\textbf {\bibinfo {volume} {93}},\ \bibinfo {pages} {034906}
  (\bibinfo {year} {2016})}\BibitemShut {NoStop}%
\bibitem [{\citenamefont {Cao}\ \emph {et~al.}(2015)\citenamefont {Cao},
  \citenamefont {Qin},\ and\ \citenamefont {Bass}}]{PhysRevC.92.024907}%
  \BibitemOpen
  \bibfield  {author} {\bibinfo {author} {\bibfnamefont {S.}~\bibnamefont
  {Cao}}, \bibinfo {author} {\bibfnamefont {G.-Y.}\ \bibnamefont {Qin}}, \ and\
  \bibinfo {author} {\bibfnamefont {S.~A.}\ \bibnamefont {Bass}},\ }\href
  {\doibase 10.1103/PhysRevC.92.024907} {\bibfield  {journal} {\bibinfo
  {journal} {Phys. Rev. C}\ }\textbf {\bibinfo {volume} {92}},\ \bibinfo
  {pages} {024907} (\bibinfo {year} {2015})}\BibitemShut {NoStop}%
\bibitem [{\citenamefont {Cao}\ \emph {et~al.}(2019)\citenamefont {Cao} \emph
  {et~al.}}]{Cao:2018ews}%
  \BibitemOpen
  \bibfield  {author} {\bibinfo {author} {\bibfnamefont {S.}~\bibnamefont
  {Cao}} \emph {et~al.},\ }\href {\doibase 10.1103/PhysRevC.99.054907}
  {\bibfield  {journal} {\bibinfo  {journal} {Phys. Rev. C}\ }\textbf {\bibinfo
  {volume} {99}},\ \bibinfo {pages} {054907} (\bibinfo {year} {2019})},\
  \Eprint {http://arxiv.org/abs/1809.07894} {arXiv:1809.07894 [nucl-th]}
  \BibitemShut {NoStop}%
\bibitem [{\citenamefont {{S.~Z.~Shi, J.~F.~Liao, and
  M.~Gyulassy}}(2018{\natexlab{a}})}]{CUJET3Arxiv18}%
  \BibitemOpen
  \bibfield  {author} {\bibinfo {author} {\bibnamefont {{S.~Z.~Shi, J.~F.~Liao,
  and M.~Gyulassy}}},\ }\href {\doibase 10.1088/1674-1137/43/4/044101}
  {\bibfield  {journal} {\bibinfo  {journal} {Chin.~Phys.~C}\ }\textbf
  {\bibinfo {volume} {43}},\ \bibinfo {pages} {044101} (\bibinfo {year}
  {2018}{\natexlab{a}})}\BibitemShut {NoStop}%
\bibitem [{\citenamefont {{S.~Z.~Shi, J.~F.~Liao, and
  M.~Gyulassy}}(2018{\natexlab{b}})}]{CUJET3CPC18}%
  \BibitemOpen
  \bibfield  {author} {\bibinfo {author} {\bibnamefont {{S.~Z.~Shi, J.~F.~Liao,
  and M.~Gyulassy}}},\ }\href {\doibase 10.1088/1674-1137/42/10/104104}
  {\bibfield  {journal} {\bibinfo  {journal} {Chin.~Phys.~C}\ }\textbf
  {\bibinfo {volume} {42}},\ \bibinfo {pages} {104104} (\bibinfo {year}
  {2018}{\natexlab{b}})}\BibitemShut {NoStop}%
\bibitem [{\citenamefont {Ke}\ \emph {et~al.}(2018)\citenamefont {Ke},
  \citenamefont {Xu},\ and\ \citenamefont {Bass}}]{PhysRevC.98.064901}%
  \BibitemOpen
  \bibfield  {author} {\bibinfo {author} {\bibfnamefont {W.}~\bibnamefont
  {Ke}}, \bibinfo {author} {\bibfnamefont {Y.}~\bibnamefont {Xu}}, \ and\
  \bibinfo {author} {\bibfnamefont {S.~A.}\ \bibnamefont {Bass}},\ }\href
  {\doibase 10.1103/PhysRevC.98.064901} {\bibfield  {journal} {\bibinfo
  {journal} {Phys. Rev. C}\ }\textbf {\bibinfo {volume} {98}},\ \bibinfo
  {pages} {064901} (\bibinfo {year} {2018})}\BibitemShut {NoStop}%
\bibitem [{\citenamefont {{S.~Li, C.~W.~Wang, X.~B.~Yuan, and
  S.~Q.~Feng}}(2018)}]{CTGUHybrid1}%
  \BibitemOpen
  \bibfield  {author} {\bibinfo {author} {\bibnamefont {{S.~Li, C.~W.~Wang,
  X.~B.~Yuan, and S.~Q.~Feng}}},\ }\href {\doibase 10.1103/PhysRevC.98.014909}
  {\bibfield  {journal} {\bibinfo  {journal} {Phys.~Rev.~C}\ }\textbf {\bibinfo
  {volume} {98}},\ \bibinfo {pages} {014909} (\bibinfo {year}
  {2018})}\BibitemShut {NoStop}%
\bibitem [{\citenamefont {Li}\ \emph {et~al.}(2019)\citenamefont {Li},
  \citenamefont {Wang}, \citenamefont {Wan},\ and\ \citenamefont
  {Liao}}]{Li:2019wri}%
  \BibitemOpen
  \bibfield  {author} {\bibinfo {author} {\bibfnamefont {S.}~\bibnamefont
  {Li}}, \bibinfo {author} {\bibfnamefont {C.}~\bibnamefont {Wang}}, \bibinfo
  {author} {\bibfnamefont {R.}~\bibnamefont {Wan}}, \ and\ \bibinfo {author}
  {\bibfnamefont {J.}~\bibnamefont {Liao}},\ }\href {\doibase
  10.1103/PhysRevC.99.054909} {\bibfield  {journal} {\bibinfo  {journal} {Phys.
  Rev.}\ }\textbf {\bibinfo {volume} {C99}},\ \bibinfo {pages} {054909}
  (\bibinfo {year} {2019})},\ \Eprint {http://arxiv.org/abs/1901.04600}
  {arXiv:1901.04600 [hep-ph]} \BibitemShut {NoStop}%
\bibitem [{\citenamefont {{J.~D.~Bjorken}}(1982)}]{bjorken1982energy}%
  \BibitemOpen
  \bibfield  {author} {\bibinfo {author} {\bibnamefont {{J.~D.~Bjorken}}},\
  }\href {http://lss.fnal.gov/archive/1982/pub/Pub-82-059-T.pdf} {\bibfield
  {journal} {\bibinfo  {journal} {Report No.FERMILAB-Pub-82/59-THY}\ }
  (\bibinfo {year} {1982})}\BibitemShut {NoStop}%
\bibitem [{\citenamefont {Thoma}\ and\ \citenamefont
  {Gyulassy}(1991)}]{Thoma:1990fm}%
  \BibitemOpen
  \bibfield  {author} {\bibinfo {author} {\bibfnamefont {M.~H.}\ \bibnamefont
  {Thoma}}\ and\ \bibinfo {author} {\bibfnamefont {M.}~\bibnamefont
  {Gyulassy}},\ }\href {\doibase 10.1016/S0550-3213(05)80031-8} {\bibfield
  {journal} {\bibinfo  {journal} {Nucl. Phys. B}\ }\textbf {\bibinfo {volume}
  {351}},\ \bibinfo {pages} {491} (\bibinfo {year} {1991})}\BibitemShut
  {NoStop}%
\bibitem [{\citenamefont {Braaten}\ and\ \citenamefont
  {Thoma}(1991{\natexlab{a}})}]{braaten1991energy}%
  \BibitemOpen
  \bibfield  {author} {\bibinfo {author} {\bibfnamefont {E.}~\bibnamefont
  {Braaten}}\ and\ \bibinfo {author} {\bibfnamefont {M.~H.}\ \bibnamefont
  {Thoma}},\ }\href {\doibase 10.1103/PhysRevD.44.1298} {\bibfield  {journal}
  {\bibinfo  {journal} {Phys. Rev. D}\ }\textbf {\bibinfo {volume} {44}},\
  \bibinfo {pages} {1298} (\bibinfo {year} {1991}{\natexlab{a}})}\BibitemShut
  {NoStop}%
\bibitem [{\citenamefont {Vija}\ and\ \citenamefont
  {Thoma}(1995)}]{Vija:1994is}%
  \BibitemOpen
  \bibfield  {author} {\bibinfo {author} {\bibfnamefont {H.}~\bibnamefont
  {Vija}}\ and\ \bibinfo {author} {\bibfnamefont {M.~H.}\ \bibnamefont
  {Thoma}},\ }\href {\doibase 10.1016/0370-2693(94)01378-P} {\bibfield
  {journal} {\bibinfo  {journal} {Phys. Lett. B}\ }\textbf {\bibinfo {volume}
  {342}},\ \bibinfo {pages} {212} (\bibinfo {year} {1995})},\ \Eprint
  {http://arxiv.org/abs/hep-ph/9409246} {arXiv:hep-ph/9409246} \BibitemShut
  {NoStop}%
\bibitem [{\citenamefont {Berrehrah}\ \emph {et~al.}(2014)\citenamefont
  {Berrehrah}, \citenamefont {Gossiaux}, \citenamefont {Aichelin},
  \citenamefont {Cassing}, \citenamefont {Torres-Rincon},\ and\ \citenamefont
  {Bratkovskaya}}]{Berrehrah:2014tva}%
  \BibitemOpen
  \bibfield  {author} {\bibinfo {author} {\bibfnamefont {H.}~\bibnamefont
  {Berrehrah}}, \bibinfo {author} {\bibfnamefont {P.~B.}\ \bibnamefont
  {Gossiaux}}, \bibinfo {author} {\bibfnamefont {J.}~\bibnamefont {Aichelin}},
  \bibinfo {author} {\bibfnamefont {W.}~\bibnamefont {Cassing}}, \bibinfo
  {author} {\bibfnamefont {J.~M.}\ \bibnamefont {Torres-Rincon}}, \ and\
  \bibinfo {author} {\bibfnamefont {E.}~\bibnamefont {Bratkovskaya}},\ }\href
  {\doibase 10.1103/PhysRevC.90.051901} {\bibfield  {journal} {\bibinfo
  {journal} {Phys. Rev. C}\ }\textbf {\bibinfo {volume} {90}},\ \bibinfo
  {pages} {051901} (\bibinfo {year} {2014})},\ \Eprint
  {http://arxiv.org/abs/1406.5322} {arXiv:1406.5322 [hep-ph]} \BibitemShut
  {NoStop}%
\bibitem [{\citenamefont {Jamal}\ and\ \citenamefont
  {Mohanty}(2021)}]{Jamal:2020emj}%
  \BibitemOpen
  \bibfield  {author} {\bibinfo {author} {\bibfnamefont {M.~Y.}\ \bibnamefont
  {Jamal}}\ and\ \bibinfo {author} {\bibfnamefont {B.}~\bibnamefont
  {Mohanty}},\ }\href {\doibase 10.1140/epjp/s13360-021-01098-4} {\bibfield
  {journal} {\bibinfo  {journal} {Eur. Phys. J. Plus}\ }\textbf {\bibinfo
  {volume} {136}},\ \bibinfo {pages} {130} (\bibinfo {year} {2021})},\ \Eprint
  {http://arxiv.org/abs/2002.09230} {arXiv:2002.09230 [nucl-th]} \BibitemShut
  {NoStop}%
\bibitem [{\citenamefont {Madni}\ \emph {et~al.}(2023)\citenamefont {Madni},
  \citenamefont {Mukherjee}, \citenamefont {Bandyopadhyay},\ and\ \citenamefont
  {Haque}}]{Madni:2022bea}%
  \BibitemOpen
  \bibfield  {author} {\bibinfo {author} {\bibfnamefont {S.}~\bibnamefont
  {Madni}}, \bibinfo {author} {\bibfnamefont {A.}~\bibnamefont {Mukherjee}},
  \bibinfo {author} {\bibfnamefont {A.}~\bibnamefont {Bandyopadhyay}}, \ and\
  \bibinfo {author} {\bibfnamefont {N.}~\bibnamefont {Haque}},\ }\href
  {\doibase 10.1016/j.physletb.2023.137714} {\bibfield  {journal} {\bibinfo
  {journal} {Phys. Lett. B}\ }\textbf {\bibinfo {volume} {838}},\ \bibinfo
  {pages} {137714} (\bibinfo {year} {2023})},\ \Eprint
  {http://arxiv.org/abs/2210.03076} {arXiv:2210.03076 [hep-ph]} \BibitemShut
  {NoStop}%
\bibitem [{\citenamefont {Peng}\ \emph {et~al.}(2024)\citenamefont {Peng},
  \citenamefont {Yu}, \citenamefont {Li}, \citenamefont {Xiong}, \citenamefont
  {Sun},\ and\ \citenamefont {Xie}}]{peng2024unraveling}%
  \BibitemOpen
  \bibfield  {author} {\bibinfo {author} {\bibfnamefont {J.}~\bibnamefont
  {Peng}}, \bibinfo {author} {\bibfnamefont {K.}~\bibnamefont {Yu}}, \bibinfo
  {author} {\bibfnamefont {S.}~\bibnamefont {Li}}, \bibinfo {author}
  {\bibfnamefont {W.}~\bibnamefont {Xiong}}, \bibinfo {author} {\bibfnamefont
  {F.}~\bibnamefont {Sun}}, \ and\ \bibinfo {author} {\bibfnamefont
  {W.}~\bibnamefont {Xie}},\ }\href {\doibase 10.1103/PhysRevD.109.096028}
  {\bibfield  {journal} {\bibinfo  {journal} {Phys. Rev. D}\ }\textbf {\bibinfo
  {volume} {109}},\ \bibinfo {pages} {096028} (\bibinfo {year} {2024})},\
  \Eprint {http://arxiv.org/abs/2401.10644} {arXiv:2401.10644 [hep-ph]}
  \BibitemShut {NoStop}%
\bibitem [{\citenamefont {Li}\ \emph {et~al.}(2021)\citenamefont {Li},
  \citenamefont {Sun}, \citenamefont {Xie},\ and\ \citenamefont
  {Xiong}}]{li2021langevin}%
  \BibitemOpen
  \bibfield  {author} {\bibinfo {author} {\bibfnamefont {S.}~\bibnamefont
  {Li}}, \bibinfo {author} {\bibfnamefont {F.}~\bibnamefont {Sun}}, \bibinfo
  {author} {\bibfnamefont {W.}~\bibnamefont {Xie}}, \ and\ \bibinfo {author}
  {\bibfnamefont {W.}~\bibnamefont {Xiong}},\ }\href {\doibase
  10.1140/epjc/s10052-021-09339-7} {\bibfield  {journal} {\bibinfo  {journal}
  {Eur. Phys. J. C}\ }\textbf {\bibinfo {volume} {81}} (\bibinfo {year}
  {2021}),\ 10.1140/epjc/s10052-021-09339-7}\BibitemShut {NoStop}%
\bibitem [{\citenamefont {{E.~Braaten and
  T.~C.~Yuan}}(1991)}]{braaten1991calculation}%
  \BibitemOpen
  \bibfield  {author} {\bibinfo {author} {\bibnamefont {{E.~Braaten and
  T.~C.~Yuan}}},\ }\href {\doibase 10.1103/PhysRevLett.66.2183} {\bibfield
  {journal} {\bibinfo  {journal} {Phys.~Rev.~Lett.}\ }\textbf {\bibinfo
  {volume} {2183}},\ \bibinfo {pages} {66} (\bibinfo {year}
  {1991})}\BibitemShut {NoStop}%
\bibitem [{\citenamefont {Braaten}\ and\ \citenamefont
  {Thoma}(1991{\natexlab{b}})}]{PhysRevD.44.R2625}%
  \BibitemOpen
  \bibfield  {author} {\bibinfo {author} {\bibfnamefont {E.}~\bibnamefont
  {Braaten}}\ and\ \bibinfo {author} {\bibfnamefont {M.~H.}\ \bibnamefont
  {Thoma}},\ }\href {\doibase 10.1103/PhysRevD.44.R2625} {\bibfield  {journal}
  {\bibinfo  {journal} {Phys. Rev. D}\ }\textbf {\bibinfo {volume} {44}},\
  \bibinfo {pages} {R2625} (\bibinfo {year} {1991}{\natexlab{b}})}\BibitemShut
  {NoStop}%
\bibitem [{\citenamefont {Blaizot}\ and\ \citenamefont
  {Iancu}(2002)}]{JeanPR02}%
  \BibitemOpen
  \bibfield  {author} {\bibinfo {author} {\bibfnamefont {J.-P.}\ \bibnamefont
  {Blaizot}}\ and\ \bibinfo {author} {\bibfnamefont {E.}~\bibnamefont
  {Iancu}},\ }\href {\doibase 10.1016/S0370-1573(01)00061-8} {\bibfield
  {journal} {\bibinfo  {journal} {Phys. Rept.}\ }\textbf {\bibinfo {volume}
  {359}},\ \bibinfo {pages} {355} (\bibinfo {year} {2002})},\ \Eprint
  {http://arxiv.org/abs/hep-ph/0101103} {arXiv:hep-ph/0101103} \BibitemShut
  {NoStop}%
\bibitem [{\citenamefont {Romatschke}\ and\ \citenamefont
  {Strickland}(2005)}]{Romatschke:2004au}%
  \BibitemOpen
  \bibfield  {author} {\bibinfo {author} {\bibfnamefont {P.}~\bibnamefont
  {Romatschke}}\ and\ \bibinfo {author} {\bibfnamefont {M.}~\bibnamefont
  {Strickland}},\ }\href {\doibase 10.1103/PhysRevD.71.125008} {\bibfield
  {journal} {\bibinfo  {journal} {Phys. Rev. D}\ }\textbf {\bibinfo {volume}
  {71}},\ \bibinfo {pages} {125008} (\bibinfo {year} {2005})},\ \Eprint
  {http://arxiv.org/abs/hep-ph/0408275} {arXiv:hep-ph/0408275} \BibitemShut
  {NoStop}%
\bibitem [{\citenamefont {{P.~B.~Gossiaux and J.~Aichelin}}(2008)}]{PBGPRC08}%
  \BibitemOpen
  \bibfield  {author} {\bibinfo {author} {\bibnamefont {{P.~B.~Gossiaux and
  J.~Aichelin}}},\ }\href {\doibase 10.1103/PhysRevC.78.014904} {\bibfield
  {journal} {\bibinfo  {journal} {Phys.~Rev.~C}\ }\textbf {\bibinfo {volume}
  {78}},\ \bibinfo {pages} {014904} (\bibinfo {year} {2008})}\BibitemShut
  {NoStop}%
\bibitem [{\citenamefont {{S.~Li and C.~W.~Wang}}(2018)}]{CTGUHybrid2}%
  \BibitemOpen
  \bibfield  {author} {\bibinfo {author} {\bibnamefont {{S.~Li and
  C.~W.~Wang}}},\ }\href {\doibase 10.1103/PhysRevC.98.034914} {\bibfield
  {journal} {\bibinfo  {journal} {Phys.~Rev.~C}\ }\textbf {\bibinfo {volume}
  {98}},\ \bibinfo {pages} {034914} (\bibinfo {year} {2018})}\BibitemShut
  {NoStop}%
\bibitem [{\citenamefont {{S.~Li, W.~Xiong, and
  R.~Z.~Wan}}(2020)}]{CTGUHybrid5}%
  \BibitemOpen
  \bibfield  {author} {\bibinfo {author} {\bibnamefont {{S.~Li, W.~Xiong, and
  R.~Z.~Wan}}},\ }\href {\doibase 10.1140/epjc/s10052-020-08708-y} {\bibfield
  {journal} {\bibinfo  {journal} {Eur. Phys. J. C}\ }\textbf {\bibinfo {volume}
  {80}},\ \bibinfo {pages} {1113} (\bibinfo {year} {2020})}\BibitemShut
  {NoStop}%
\bibitem [{\citenamefont {Li}\ and\ \citenamefont {Liao}(2020)}]{Li:2019lex}%
  \BibitemOpen
  \bibfield  {author} {\bibinfo {author} {\bibfnamefont {S.}~\bibnamefont
  {Li}}\ and\ \bibinfo {author} {\bibfnamefont {J.}~\bibnamefont {Liao}},\
  }\href {\doibase 10.1140/epjc/s10052-020-8243-9} {\bibfield  {journal}
  {\bibinfo  {journal} {Eur. Phys. J. C}\ }\textbf {\bibinfo {volume} {80}},\
  \bibinfo {pages} {671} (\bibinfo {year} {2020})},\ \Eprint
  {http://arxiv.org/abs/1912.08965} {arXiv:1912.08965 [hep-ph]} \BibitemShut
  {NoStop}%
\bibitem [{\citenamefont {Weldon}(1983)}]{Weldon:1983jn}%
  \BibitemOpen
  \bibfield  {author} {\bibinfo {author} {\bibfnamefont {H.~A.}\ \bibnamefont
  {Weldon}},\ }\href {\doibase 10.1103/PhysRevD.28.2007} {\bibfield  {journal}
  {\bibinfo  {journal} {Phys. Rev. D}\ }\textbf {\bibinfo {volume} {28}},\
  \bibinfo {pages} {2007} (\bibinfo {year} {1983})}\BibitemShut {NoStop}%
\bibitem [{\citenamefont {Xu}\ and\ \citenamefont
  {Greiner}(2005)}]{PhysRevC.71.064901}%
  \BibitemOpen
  \bibfield  {author} {\bibinfo {author} {\bibfnamefont {Z.}~\bibnamefont
  {Xu}}\ and\ \bibinfo {author} {\bibfnamefont {C.}~\bibnamefont {Greiner}},\
  }\href {\doibase 10.1103/PhysRevC.71.064901} {\bibfield  {journal} {\bibinfo
  {journal} {Phys. Rev. C}\ }\textbf {\bibinfo {volume} {71}},\ \bibinfo
  {pages} {064901} (\bibinfo {year} {2005})}\BibitemShut {NoStop}%
\bibitem [{\citenamefont {Wong}(1996)}]{PhysRevC.54.2588}%
  \BibitemOpen
  \bibfield  {author} {\bibinfo {author} {\bibfnamefont {S.~M.~H.}\
  \bibnamefont {Wong}},\ }\href {\doibase 10.1103/PhysRevC.54.2588} {\bibfield
  {journal} {\bibinfo  {journal} {Phys. Rev. C}\ }\textbf {\bibinfo {volume}
  {54}},\ \bibinfo {pages} {2588} (\bibinfo {year} {1996})}\BibitemShut
  {NoStop}%
\bibitem [{\citenamefont {{B.~L.~Combridge}}(1979)}]{Combridge79}%
  \BibitemOpen
  \bibfield  {author} {\bibinfo {author} {\bibnamefont {{B.~L.~Combridge}}},\
  }\href {\doibase 10.1016/0550-3213(79)90449-8} {\bibfield  {journal}
  {\bibinfo  {journal} {Nucl.~Phys.~B}\ }\textbf {\bibinfo {volume} {151}},\
  \bibinfo {pages} {429} (\bibinfo {year} {1979})}\BibitemShut {NoStop}%
\bibitem [{\citenamefont {Peign\'e}\ and\ \citenamefont
  {Peshier}(2008{\natexlab{a}})}]{peigne2008collisional}%
  \BibitemOpen
  \bibfield  {author} {\bibinfo {author} {\bibfnamefont {S.}~\bibnamefont
  {Peign\'e}}\ and\ \bibinfo {author} {\bibfnamefont {A.}~\bibnamefont
  {Peshier}},\ }\href {\doibase 10.1103/PhysRevD.77.014015} {\bibfield
  {journal} {\bibinfo  {journal} {Phys. Rev. D}\ }\textbf {\bibinfo {volume}
  {77}},\ \bibinfo {pages} {014015} (\bibinfo {year}
  {2008}{\natexlab{a}})}\BibitemShut {NoStop}%
\bibitem [{\citenamefont {Peign\'e}\ and\ \citenamefont
  {Peshier}(2008{\natexlab{b}})}]{PhysRevD.77.114017}%
  \BibitemOpen
  \bibfield  {author} {\bibinfo {author} {\bibfnamefont {S.}~\bibnamefont
  {Peign\'e}}\ and\ \bibinfo {author} {\bibfnamefont {A.}~\bibnamefont
  {Peshier}},\ }\href {\doibase 10.1103/PhysRevD.77.114017} {\bibfield
  {journal} {\bibinfo  {journal} {Phys. Rev. D}\ }\textbf {\bibinfo {volume}
  {77}},\ \bibinfo {pages} {114017} (\bibinfo {year}
  {2008}{\natexlab{b}})}\BibitemShut {NoStop}%
\bibitem [{\citenamefont {Bellwied}\ \emph {et~al.}(2015)\citenamefont
  {Bellwied}, \citenamefont {Borsanyi}, \citenamefont {Fodor}, \citenamefont
  {G{\"u}nther}, \citenamefont {Katz}, \citenamefont {Ratti},\ and\
  \citenamefont {Szabo}}]{Bellwied:2015rza}%
  \BibitemOpen
  \bibfield  {author} {\bibinfo {author} {\bibfnamefont {R.}~\bibnamefont
  {Bellwied}}, \bibinfo {author} {\bibfnamefont {S.}~\bibnamefont {Borsanyi}},
  \bibinfo {author} {\bibfnamefont {Z.}~\bibnamefont {Fodor}}, \bibinfo
  {author} {\bibfnamefont {J.}~\bibnamefont {G{\"u}nther}}, \bibinfo {author}
  {\bibfnamefont {S.~D.}\ \bibnamefont {Katz}}, \bibinfo {author}
  {\bibfnamefont {C.}~\bibnamefont {Ratti}}, \ and\ \bibinfo {author}
  {\bibfnamefont {K.~K.}\ \bibnamefont {Szabo}},\ }\href {\doibase
  10.1016/j.physletb.2015.11.011} {\bibfield  {journal} {\bibinfo  {journal}
  {Phys. Lett. B}\ }\textbf {\bibinfo {volume} {751}},\ \bibinfo {pages} {559}
  (\bibinfo {year} {2015})},\ \Eprint {http://arxiv.org/abs/1507.07510}
  {arXiv:1507.07510 [hep-lat]} \BibitemShut {NoStop}%
\bibitem [{\citenamefont {Borsanyi}\ \emph {et~al.}(2020)\citenamefont
  {Borsanyi}, \citenamefont {Fodor}, \citenamefont {Guenther}, \citenamefont
  {Kara}, \citenamefont {Katz}, \citenamefont {Parotto}, \citenamefont
  {Pasztor}, \citenamefont {Ratti},\ and\ \citenamefont
  {Szabo}}]{Borsanyi:2020fev}%
  \BibitemOpen
  \bibfield  {author} {\bibinfo {author} {\bibfnamefont {S.}~\bibnamefont
  {Borsanyi}}, \bibinfo {author} {\bibfnamefont {Z.}~\bibnamefont {Fodor}},
  \bibinfo {author} {\bibfnamefont {J.~N.}\ \bibnamefont {Guenther}}, \bibinfo
  {author} {\bibfnamefont {R.}~\bibnamefont {Kara}}, \bibinfo {author}
  {\bibfnamefont {S.~D.}\ \bibnamefont {Katz}}, \bibinfo {author}
  {\bibfnamefont {P.}~\bibnamefont {Parotto}}, \bibinfo {author} {\bibfnamefont
  {A.}~\bibnamefont {Pasztor}}, \bibinfo {author} {\bibfnamefont
  {C.}~\bibnamefont {Ratti}}, \ and\ \bibinfo {author} {\bibfnamefont {K.~K.}\
  \bibnamefont {Szabo}},\ }\href {\doibase 10.1103/PhysRevLett.125.052001}
  {\bibfield  {journal} {\bibinfo  {journal} {Phys. Rev. Lett.}\ }\textbf
  {\bibinfo {volume} {125}},\ \bibinfo {pages} {052001} (\bibinfo {year}
  {2020})},\ \Eprint {http://arxiv.org/abs/2002.02821} {arXiv:2002.02821
  [hep-lat]} \BibitemShut {NoStop}%
\end{thebibliography}
